\documentclass[reprint,amsmath, amsmath,amssymb,
 aps,nofootinbib,amssymb,aps,prd,onecolumn]{revtex4-1} 
\usepackage{soul}
\usepackage{hyperref}
\usepackage{float} 
 \usepackage{caption}
 \usepackage{subcaption}
\usepackage{graphicx}
\usepackage{dblfloatfix}
\usepackage{placeins}
\usepackage{xcolor}
\usepackage{slashed}
\usepackage{graphicx}% Include figure files
\usepackage{dcolumn}% Align table columns on decimal point
\usepackage{bm}% bold math
\begin{document}

%\preprint{APS/123-QED}

\title{Neutrino Mass, Vacuum Stability and Higgs Inflation\\
with Vector-Like Quarks and a Single Right-Handed Neutrino}% Force line breaks with \\

\author{Canan Karahan}
 \email{ckarahan@itu.edu.tr}%Lines break automatically or can be forced with \\
 \affiliation{%
National Defence University, Turkish Naval Academy, Department of Basic Sciences, 34942 Tuzla, {\.I}stanbul, T\"{u}rkiye
}

\begin{abstract}
We investigate a Standard Model extension containing $n$ degenerate down-type isosinglet vector-like quarks (VLQs) with masses $M_{\mathcal D}$ and Yukawa couplings $y_{\mathcal D}$, supplemented by a single right-handed neutrino (RHN), aiming to simultaneously address neutrino mass generation, electroweak vacuum stability, and Higgs inflation. The VLQs play the dominant role in stabilizing the Higgs potential through their impact on the renormalization-group evolution, while the RHN generates light neutrino masses via a Type-I seesaw mechanism and smooths the high-scale running of the Higgs quartic coupling in the inflationary regime. We perform a two-loop Standard Model renormalization-group equation analysis supplemented by the one-loop contributions of the VLQs and the RHN, with proper matching across their mass thresholds. Using these RG trajectories, we identify the regions in $(n,\, y_{\mathcal D},\, M_{\mathcal D})$ that stabilize the Higgs potential up to the Planck scale while satisfying experimental constraints. Employing the RG-improved Higgs potential in the metric formulation of non-minimal Higgs inflation, we compute the inflationary observables $n_s$ and $r$. The SM+$(n)$VLQ+RHN framework yields predictions consistent with the latest Planck-LB-BK18 and ACT-LB-BK18 data, while simultaneously ensuring electroweak vacuum stability and phenomenologically viable neutrino masses within well-defined regions of parameter space. For comparison, we also investigate the SM+$(n)$VLQ limit and present its vacuum
stability and Higgs inflation predictions as a reference to quantify the stabilizing role of the VLQ sector.
\end{abstract}

\pacs{Valid PACS appear here}% PACS, the Physics and Astronomy
\maketitle

%\tableofcontents

\section{Introduction}

The Standard Model (SM) of particle physics has been remarkably successful in describing a wide range of phenomena. Nevertheless, several experimental observations and theoretical considerations—most notably neutrino masses \cite{SNO2001, SuperK1998}, the metastability of the electroweak vacuum \cite{Degrassi2012, Buttazzo2013}, and the need for a viable mechanism for cosmic inflation \cite{Starobinsky1980, Bezrukov2008}—point to the need for extensions of the Standard Model. In the following, we briefly discuss these open issues and their implications
for physics beyond the Standard Model.

First, neutrino oscillation experiments have established that neutrinos are not massless, 
contrary to the SM prediction, but have tiny yet finite masses on the sub-eV scale 
\cite{SNO2001,SuperK1998}. 
This calls for an extension of the SM to generate neutrino masses. 
The most straightforward explanation is the Type-I seesaw mechanism 
\cite{Minkowski1977,GellMann1979,Yanagida1979,MohapatraSenjanovic1980}, 
which introduces heavy right-handed neutrinos (singlet under SM gauge interactions) 
that couple to SM lepton doublets.
Beside the Type-I seesaw mechanism, there are Type-II \cite{Mohapatra:1981, Cheng:1980}, Type-III \cite{Foot:1989} and Type-3/2 \cite{Demir:2021} seesaw mechanisms, which include a $SU(2)_L$ triplet scalar field, a $SU(2)_L$ triplet fermion field and spin-3/2 field, respectively. Among these possibilities, the Type-I seesaw mechanism provides a simple and
well-motivated framework for generating light neutrino masses.

Second, the stability of the electroweak (EW) vacuum poses a theoretical puzzle.
Given the measured Higgs boson mass $m_h \approx 125~\text{GeV}$ and top-quark
mass $m_t \approx 172.56~\text{GeV}$, the RG evolution of the SM
couplings to high energies indicates that the Higgs self-coupling $\lambda_H$
runs to negative values at a scale of order $10^{10}~\text{GeV}$. In other words, the Higgs potential $V(H)$ develops a second, deeper minimum at high field values, rendering the EW vacuum metastable (with a lifetime longer than the age of the Universe, but not absolutely stable) \cite{Sher1989, Degrassi2012, Buttazzo2013}. This theoretically challenging situation — often called the vacuum stability problem — has been confirmed by high-precision  2-loop \cite{Degrassi2012} and 3-loop \cite{Chetyrkin:2012rz} computations of the Higgs effective potential. 
In literature, numerous extensions of the Standard Model have been proposed to address this issue, such as introducing additional scalar fields to raise the Higgs quartic coupling at high energies \cite{Gonderinger:0910.3167, Ghorbani:2104.09542, Peli:2022, Hiller:2024}, or new fermions/gauge sectors that alter the renormalization-group (RG) running of $\lambda_H$ \cite{Rodejohann:2012, Datta:2013, Oda:2015}. In particular, heavy vector-like quarks (VLQs) are an attractive possibility \cite{Gopalakrishna:2019, Arsenault:2023vlf}: being color-triplet fermions that receive gauge-invariant mass terms, they are free of gauge anomalies and often appear in extensions like extra-dimensional models or composite Higgs scenarios. A down-type (up-type) isosinglet VLQ (often denoted $\mathcal{D}$ ($\mathcal{U}$) ) carries the same SM gauge quantum numbers as a right-handed bottom (top) quark (charge $-1/3$ ($+2/3$), singlet under $SU(2)_L$) and can mix with the third-generation quarks  via its Yukawa interactions. The presence of such a VLQ can modify the RG evolution of $\lambda_H$. Notably, if the VLQ has a mass of order the TeV scale and a moderate mixing with the third-generation quarks, it can reduce the effective top Yukawa coupling and strong coupling at high energies, thereby delaying or preventing the eventual turn-over of $\lambda_H$ to negative values. Indeed, recent studies have shown that the addition of  vector-like quarks can stabilize the Higgs potential up to the Planck scale for appropriate parameter choices \cite{Gopalakrishna:2019, Arsenault:2023vlf}, while also providing well-motivated links to gauge coupling unification, Grand Unified Theory (GUT) embeddings, and successful Higgs inflation scenarios \cite{referee_1,referee_2}.

Third, the origin of the cosmic inflationary epoch remains an open question at the intersection of particle physics and cosmology. The simplest models of cosmic inflation postulate a new scalar (inflaton) field to drive exponential expansion in the early Universe \cite{Lerner:2009xg, Nakayama:2010}  inflation \cite{Barvinsky:2008, Bauer:2008, Rubio:2018}: the identification of the SM Higgs field itself as the inflaton. Higgs field is coupled non-minimally to gravity via a term $\xi H^\dagger H R$ (where $R$ is the Ricci scalar curvature and $\xi$ is a dimensionless coupling). For $\xi \sim 10^4$, the Higgs potential in the Einstein frame becomes sufficiently flat at large field values to sustain slow-roll inflation, yielding predictions for the spectral index $n_s$ and tensor-to-scalar ratio $r$ in excellent agreement with observations. Higgs inflation is highly economical – it requires no new dynamical degrees of freedom beyond the SM – but its consistency hinges on the stability of the Higgs potential up to the inflationary scales (typically just below the Planck scale). If the potential has a destabilizing turn ($\lambda_H<0$) at field values below the inflationary plateau, the Higgs field could tunnel to the lower vacuum instead of inflating, ruining the scenario. Ensuring vacuum stability is therefore essential for Higgs inflation to be viable. For the current central values of the Higgs boson mass $m_h$  and top-quark mass $m_t$,
the Standard Model alone does not achieve absolute electroweak vacuum stability. This provides strong motivation to consider new physics that can stabilize the Higgs potential and simultaneously allow the Higgs to act as the inflaton.

In this paper, we investigate a minimal phenomenological extension of the SM that addresses all three issues discussed above – neutrino masses, electroweak vacuum stability, and Higgs inflation – in a unified framework. The model introduces just two new sectors of fermionic states: (i) a set of n degenerate isosinglet down-type VLQs, and (ii) a singlet RHN. The RHN, endowed with a large Majorana mass $M_N$, generates light neutrino masses through the Type-I seesaw mechanism and, through its Yukawa coupling,
modifies the RG evolution of the Higgs quartic coupling in the inflationary regime,
thereby impacting Higgs inflation \footnote{In this work, the RHN is primarily introduced to generate neutrino masses and to affect the renormalization-group evolution of the Higgs sector. A realistic neutrino oscillation pattern would require at least two RHNs, which can be straightforwardly incorporated without altering the qualitative features of our framework. A brief analysis and discussion regarding this is provided in Appendix \ref{sec:RHNs}}The down-type VLQs  interact with the SM through the color and hypercharge gauge couplings and via a Yukawa coupling $y_\mathcal{D}$ to the Higgs doublet. For appropriate values of the VLQ mass $M_\mathcal{D}$ and the Yukawa coupling $y_\mathcal{D}$, these interactions modify the renormalization-group running of the Higgs quartic coupling $\lambda_H$, allowing the Higgs potential to remain stable up to the Planck scale.
  Importantly, this improved stability of the Higgs sector enables successful Higgs inflation: with the addition of RHN and a set of n degenerate VLQs, the Higgs field can have a stable inflationary potential when coupled to gravity with a non-minimal coupling $\xi$, as per the usual Higgs inflation paradigm. In other words, the model simultaneously allows neutrino mass generation (via the seesaw mechanism) and Higgs-driven inflation, while stabilizing the electroweak vacuum. This kind of two-particle type extension of the SM represents a phenomenologically minimalistic yet powerful approach to tackle multiple problems at once, and it remains consistent with current experimental constraints. We emphasize that both new fermions introduced here are singlets under $SU(2)_L$ and thus do not spoil the $\rho$-parameter (custodial symmetry) at tree-level; moreover, being gauge singlets or vector-like, they do not introduce anomalies.

The paper is organized as follows. In Sec.\ref{sec:model}, we define the particle content and the Lagrangian of the models under consideration, including a detailed description of the Yukawa sector involving the VLQs and the RHN, and we also discuss the relevant parameter space together with the theoretical and experimental constraints imposed on the model.  In Sec.\ref{sec:vacuum}, we investigate the vacuum stability in the presence
of the new fermionic degrees of freedom by studying the RG
evolution of the Higgs quartic coupling, and we identify the regions of
parameter space in which the Higgs potential remains stable up to the Planck
scale.
In Sec.\ref{higgs-inf}, we study Higgs inflation based on the RG-improved Higgs potential,
where the non-minimal coupling $\xi$ is fixed by the observed
amplitude of scalar perturbations and the
resulting inflationary predictions are compared with cosmological data.
Finally, our conclusions are presented in Sec.\ref{conc}. 
\section{Model Framework}
\label{sec:model}
We extend the SM by introducing two additional sectors:
(i) a set of n degenerate down-type isosinglet VLQs, and
(ii) a single Majorana RHN.
The resulting framework, denoted as $SM+(n)VLQ+RHN$, is designed to address three long-standing shortcomings of the SM in a unified manner: the generation of neutrino masses, the stabilization of the electroweak vacuum, and the realization of successful Higgs inflation. For comparison, we also analyze the $SM+(n)VLQ$ limit obtained by removing the RHN sector, thereby isolating the stabilizing role of the VLQs while excluding the destabilizing contribution of the RHN.

\subsection{Particle Content and Gauge Representations}

The new fields added to the SM are the following:

\begin{itemize}
    \item Down-type isosinglet vector-like quarks
    \begin{equation}
        D_{L,R}^{(i)} \sim (3,\,1,\,-1/3), \qquad i = 1,\ldots,n,
    \end{equation}
    transforming as color triplets, electroweak singlets, and hypercharge $Y=-1/3$;

    \item Majorana right-handed neutrino
    \begin{equation}
        N \sim (1,\,1,\,0),
    \end{equation}
    which is fully neutral under the SM gauge group.
\end{itemize}

While we focus on down-type isosinglet VLQs, it is important to note that up-type counterparts ($U \sim (3, 1, 2/3)$) are also extensively discussed in the literature regarding Higgs stability and inflation. However, the choice of down-type VLQs in this work is motivated by their more conservative impact on the high-scale validity of the theory. Specifically, an up-type VLQ contributes $\Delta \beta_{g_1} = \left(16/15\right) g_1^3$ to the hypercharge beta function, which is four times the contribution of a down-type singlet ($\Delta \beta_{g_1} = \left(4/15\right) g_1^3$).  A straightforward numerical evaluation of the one-loop RGEs reveals that for the high number of VLQs ($n \le 10$) explored in our analysis, the up-type sector accelerates the running of $g_1$ so drastically that the theory is driven toward a Landau pole well before reaching the Planck scale. Such rapid loss of perturbativity in models with large representations of vector-like fermions is a well-known theoretical constraint \cite{Adhikary:2024}. Consequently, down-type VLQs offer a more extended perturbative parameter space, allowing for a consistent evaluation of the RG-improved potential during Higgs inflation."

\subsection{Lagrangian Structure}

\subsubsection{Down-Type Isosinglet Vector-Like Quark Sector
}
The kinetic and mass terms of down-type isosinglet VLQ sector are
\begin{equation}
    \mathcal{L}_{\mathcal{D}}= 
    \overline{\mathcal{D}} \, i \slashed{D} \, \mathcal{D} 
    - M_{\mathcal{D}} \, \overline{\mathcal{D}} \mathcal{D},
\end{equation}
where $M_{\mathcal{D}}$ is a gauge-invariant Dirac mass. For an $SU(2)_L$ singlet with hypercharge $Y=-\tfrac{1}{3}$, the covariant derivative is
$D_\mu = \partial_\mu - i g_3 T^a G_\mu^a - i g' Y B_\mu$.

Gauge invariance permits for a single renormalizable Yukawa operator:
\begin{equation}
    \mathcal{L}_{\mathcal{D}}^{\rm Yuk} =
    - y_\mathcal{D} \, \overline{Q}_L \, H \, \mathcal{D}_R + \text{h.c.},
    \label{eq:Yuk_D}
\end{equation}
where $Q_L$ is the SM quark doublet and $H$ is the usual Higgs doublet. In the following, we assume that the down-type VLQs couple only to the third-generation Standard Model quarks. 
This choice suppresses potentially dangerous flavor-changing neutral currents and is consistent with current experimental bounds. For RG evolution we assume all VLQs to be mass and Yukawa coupling degenerate:
\begin{equation}
    M_{\mathcal{D}}^{(i)} = M_{\mathcal{D}}, \qquad y_{\mathcal{D}}^{(i)} = y_{\mathcal{D}}.
\end{equation}

\subsubsection{Right-Handed Neutrino Sector}
The RHN kinetic and Majorana mass terms are
\begin{equation}
    \mathcal{L}_{N} =
    \frac{1}{2} \, \overline{N} \, i \slashed{\partial} \, N
    - \frac{1}{2} M_N \, \overline{N^c} N .
\end{equation}
The neutrino Yukawa interaction is
\begin{equation}
    \mathcal{L}_{N}^{\rm Yuk} =
    - y_N \, \overline{L} \, \widetilde{H} \, N + \text{h.c.},
\end{equation}
producing a Dirac mass $m_D = y_N v/\sqrt{2}$ after electroweak symmetry breaking. Here, $\tilde H \equiv i \sigma_2 H^*$ denotes the $SU(2)_L$-conjugate Higgs doublet.

A light neutrino mass arises from the Type-I seesaw mechanism:
\begin{equation}
\label{seesaw}
    m_\nu = \frac{ y_N^2 v^2}{2M_N}.
\end{equation}

\subsection{Parameter Space and Model Limits}

The full parameter set of the extended model is
\begin{equation}
    \big\{ n,\; M_{\mathcal{D}},\; y_{\mathcal{D}},\; M_N,\; y_N \big\}.
\end{equation}

To reproduce the characteristic light–neutrino mass scale 
$m_\nu \simeq 0.05~\text{eV}$ indicated by oscillation data 
\cite{SuperK1998,PDG2023},  we fix the
right-handed neutrino parameters to the values\footnote{These values are not only consistent with neutrino oscillation data,
they are also chosen such that the RHN sector plays an active role in shaping the RG evolution of the Higgs quartic coupling.}
\begin{equation}
    M_N \simeq 10^{14} \, {\rm GeV},
    \qquad 
    y_N \simeq 0.42 ,
\end{equation}

which yield the correct order of magnitude for the Type-I seesaw relation  Eq.(\ref{seesaw}). With the neutrino sector thus fixed, the remaining free parameters of the model reduce to the three–dimensional set 
\(\{\, n,\; M_{\mathcal{D}},\; y_{\mathcal{D}} \,\}\), which fully determines the vacuum stability and Higgs–inflation phenomenology studied in the following sections. An upper bound $n \leq 10$ is imposed by the requirement of perturbativity of the strong gauge coupling.  Beyond this value, the cumulative contribution of the VLQs to the QCD beta function drives $g_3$ into the non-perturbative regime at scales relevant for vacuum stability and Higgs inflation. From this point onward, we present our analysis under two model frameworks: the full $SM+(n)VLQ+RHN$ scenario and its $SM+(n)VLQ$ limit, in which the right-handed neutrino sector is removed.\\

\subsection{Experimental and Theoretical limits}
\subsubsection{Experimental constraints on the Yukawa coupling $y_{\mathcal{D}}$ and mass $M_{\mathcal{D}}$}
In the SM extended by  down-type isosinglet VLQs  with the mass $M_{\mathcal{D}}$ and Yukawa interaction $y_{\mathcal{D}}$, the mass terms after electroweak symmetry breaking lead to a left-handed mixing angle defined by
\begin{equation}
\label{theta}
\sin\theta_{L} \;\simeq\; \frac{y_D\,v}{\sqrt{2}\,M_D},
\end{equation}

where $v \simeq 246~{\rm GeV}$ is the Higgs vacuum expectation value.  
This relation follows from the standard diagonalization of the VLQ–SM mass matrix and is consistent with the formalism developed in vector-like quark studies \cite{AguilarSaavedra:2013qpa, Lavoura:1992np}.

Current LHC searches set stringent lower bounds on the mass of down-type 
singlet VLQs.  ATLAS Run~2 pair-production analyses exclude 
$M_D \lesssim 1.3~\text{TeV}$ in the singlet-$\mathcal{D}$ ~\cite{ATLAS:VLQpair2018}, 
while the most recent CMS search at $\sqrt{s}=13$~TeV with 
$138~\text{fb}^{-1}$ further strengthens this limit by excluding 
$M_D \lesssim 1.5 ~\text{TeV}$ at 95\% C.L. 
~\cite{CMS:2209.07327}.  
These results, representative of the latest ATLAS and CMS constraints on 
pair-produced singlet VLQs, consistently indicate that masses below the 
$\sim 1.5~\text{TeV}$ range are ruled out.  

In addition to the mass bounds, the mixing of a down-type singlet vector-like quark with the SM bottom quark is strongly constrained by electroweak precision data. Such mixing induces a tree-level modification of the $Z\bar b b$ couplings, approximately given by $\delta g_L^b \simeq -\theta_L^2$. The excellent agreement between the LEP/SLC measurements and the Standard Model predictions, therefore, allows only tiny deviations. Global fits to the Z-pole observables consequently constrain the left-handed mixing angle to the percent level, $|\sin\theta_L| \lesssim \mathcal{O}(10^{-2})$, almost independently of the VLQ mass. This conclusion has been consistently confirmed in modern analyses of VLQ mixing \cite{Chen:2017qev, AguilarSaavedra:2013qpa, Lavoura:1992np}. 

In light of these constraints, and to ensure full compatibility with current
experimental data, we adopt a conservative benchmark setup for our numerical
analysis. In particular, the VLQ Yukawa coupling $y_{\mathcal{D}}$ is considered
within a range such that the induced left-handed mixing angle remains in the
experimentally allowed regime at the level of $\mathcal{O}(10^{-2})$, consistent
with bounds from electroweak precision observables. Furthermore, we impose a
lower bound on the VLQ mass, $M_D \gtrsim 1.5~\text{TeV}$, consistent with current
LHC searches.

\subsubsection{Theoretical constraints on the Yukawa coupling $y_{\mathcal{D}}$}

In addition to the experimental limits discussed above, the down-type VLQ 
Yukawa coupling $y_{\mathcal{D}}$ is subject to several theoretical consistency 
conditions. These constraints ensure that the theory remains perturbative, 
unitary, and free of Landau poles up to the Planck scale.\\

\paragraph{Perturbativity and Landau–pole constraints}

The Yukawa interaction Eq.(\ref{eq:Yuk_D}) of the down–type 
VLQ is subject to theoretical consistency requirements that 
limit its viable parameter range.  These conditions ensure the perturbative 
validity of the theory and prevent the appearance of ultraviolet Landau 
singularities.

The perturbative expansion demands
\begin{equation}
|y_{\mathcal{D}}| \lesssim \sqrt{4\pi}.
\end{equation}
In practice, phenomenological analyses of VLQ models commonly impose a 
stronger requirement, $|y_{\mathcal{D}}|\lesssim 1$, to avoid rapid growth of the cubic 
term in the beta function,
\[
\beta_{y_D} \sim \frac{y_{\mathcal{D}}^3}{16\pi^2},
\]
which is known to dominate the running at large Yukawa couplings.  
This bound ensures that perturbativity is maintained over the entire range of 
scales relevant for our RGE evolution.

Above the VLQ mass threshold, $\mu \simeq M_D = \mathcal{O}({\rm TeV})$, 
the Yukawa coupling becomes active in the renormalization-group evolution and 
starts to run according to the full SM\(+n\)VLQ+RHN beta function.  
The growth is accelerated by the cubic $y_D^3$ term, and sufficiently large 
values of $y_D$ at the matching scale can drive the coupling to a Landau pole 
far below the Planck scale.  
Requiring the absence of a Landau pole below the Planck scale and the 
validity of perturbation theory up to the scales relevant for electroweak 
vacuum stability and Higgs inflation therefore imposes a stringent upper bound 
on the VLQ Yukawa coupling.  
To remain safely within the perturbative regime, we therefore adopt the following conservative benchmark
\begin{equation}
|y_{\mathcal{D}}| \lesssim 0.5,
\end{equation}
where the precise bound depends on the number of VLQs and the scale up to which 
perturbativity is required.  
Larger values typically lead to a rapid loss of perturbative control once 
renormalization-group effects are taken into account~\cite{Adhikary:2024}.

These constraints guarantee that the Yukawa sector remains under perturbative 
control from the VLQ threshold up to the ultraviolet scales probed in our 
vacuum-stability and Higgs-inflation analysis.\\

\paragraph{Tree-level unitarity}
Perturbative unitarity of the $2\!\to\!2$ scattering amplitudes places an
additional theoretical bound on the eigenvalues of the Yukawa interaction
matrix. Adopting the standard partial-wave unitarity condition for
$s$-wave amplitudes, as commonly employed in the literature ~\cite{DicusMathur1973,Chanowitz1978}, we require that
all such amplitudes satisfy the condition $|a_0| < 1$, which implies,  
for a single down-type VLQ,
\begin{equation}
|y_{\mathcal{D}}| \lesssim \sqrt{8\pi}
\end{equation}
which is considerably weaker than the perturbativity and Landau pole limits 
discussed above.  
This bound nonetheless defines the absolute perturbative domain of the theory.

The vacuum-stability requirement, in contrast, yields a far stronger restriction 
on $y_{\mathcal{D}}$ through its renormalization–group effects on the Higgs quartic 
coupling.  
Since vacuum stability constitutes a central part of our analysis, the full 
RGE-based stability conditions—together with the experimental and theoretical 
constraints discussed above—are examined in detail in the following section.

\section{Vacuum Stability Analysis}
\label{sec:vacuum}
The scale dependence of the Higgs quartic coupling $\lambda(\mu)$ is governed by  the RGEs presented in Appendix~\ref{sec:RGE_1}. For numerical integration, all SM couplings are initialized at the top–quark mass scale 
$\,\mu = m_t\,$ using the input values listed in Appendix~\ref{sec:InputValues}.
In the SM, the large top Yukawa coupling drives $\lambda(\mu)$ negative around 
$\mu \sim 10^{10}\,\text{GeV}$ when the Higgs and top-quark masses are set to their 
current central values, $m_h \simeq 125~\text{GeV}$ and $m_t \simeq 172.56~\text{GeV}$ 
\cite{PDG2023,Buttazzo2013}.
Both the down-type vector-like quark (VLQ) sector and the right-handed neutrino (RHN)  affect this running: VLQs contribute \emph{positively} to $\beta_\lambda$, while the RHN 
contributes \emph{negatively}. 
The vacuum stability of the model therefore depends sensitively on the VLQ Yukawa 
coupling $y_{\mathcal{D}}$, the number of VLQs $n$, and the VLQ mass threshold $M_{\mathcal{D}}$.

\subsection{Vacuum Stability in the $SM+(n)VLQ+RHN$ Framework } 
We solve the coupled RGEs from the electroweak scale up to the Planck scale using 
one-loop BSM contributions and two-loop SM contributions.
The effective VLQ threshold is implemented at $\mu = M_{\mathcal{D}}$, at which the matching 
conditions for gauge, Yukawa and quartic couplings are applied.
The RHN threshold is fixed at $M_N = 10^{14}\,\text{GeV}$.

First, for each parameter point $(n, y_{\mathcal{D}}, M_{\mathcal{D}})$ we evolve the RGEs up to the Planck 
scale and determine the minimum value of the running Higgs quartic coupling,
\begin{equation}
    \lambda_{\min} \equiv \min_{\mu \leq M_{\rm Pl}} \lambda(\mu).
\end{equation}
By scanning over $y_{\mathcal{D}}$ for fixed $(n, M_{\mathcal{D}})$, we identify the critical value at 
which $\lambda(\mu)$ first becomes negative, thereby determining the boundary 
beyond which vacuum stability is lost.  
Points with $\lambda_{\min} > 0$ indicate an absolutely stable electroweak vacuum, while 
$\lambda_{\min} < 0$ corresponds to a metastable or unstable vacuum, depending on whether the tunneling lifetime exceeds 
the age of the Universe.

Fig.\ref{fig:vacuum_fullmodel} illustrates the vacuum--stability properties of the SM+$(n)$VLQ+RHN framework.
Each panel shows the minimum value of the Higgs quartic coupling, $\lambda_{\min}$, as a function of the VLQ Yukawa coupling $y_{\mathcal{D}}$ for $n=1\text{--}10$ and for the benchmark masses $M_{\mathcal{D}} = 1.5,;3.0,;5.0~\text{TeV}$, respectively.
In all cases, the RHN Yukawa coupling is fixed to $y_N = 0.42$\footnote{
Throughout this work, the Yukawa couplings $y_D$ and $y_N$ denote the input values defined at their respective mass thresholds,
$y_D(M_D)$ and $y_N(M_N)$, and are evolved according to the renormalization-group equations above these scales.
}. \\

The behavior of the curves determines 
which combinations of $(n, y_{\mathcal{D}}, M_{\mathcal{D}})$ lead to a stable electroweak vacuum. The main features of Fig.\ref{fig:vacuum_fullmodel} can be summarized as follows: 
\begin{enumerate}
  \item For fixed values of $(n,M_{\mathcal{D}})$, the minimum value of the Higgs quartic coupling,
$\lambda_{\min}$, exhibits a monotonic decrease as the VLQ Yukawa coupling $y_{\mathcal{D}}$ is increased.
The point at which a given trajectory intersects the line
$\lambda_{\min}=0$ signals the onset of vacuum instability for the
corresponding $(n,M_{\mathcal{D}})$.
Parameter regions with $\lambda_{\min}>0$ correspond to an absolutely stable
electroweak vacuum, whereas $\lambda_{\min}<0$ indicates a metastable or
unstable Higgs potential.
 \item Comparing the three panels, the range of Yukawa 
  couplings $y_{\mathcal{D}}$ that keeps $\lambda_{\min}>0$ becomes more restrictive as VLQ mass 
  $M_{\mathcal{D}}$ increase.
\begin{figure*}[h]
    \centering

    %--- panel (a): MD = 1.5 TeV ---%
    \begin{minipage}[t]{0.5\textwidth}
        \centering
        \includegraphics[width=\linewidth]{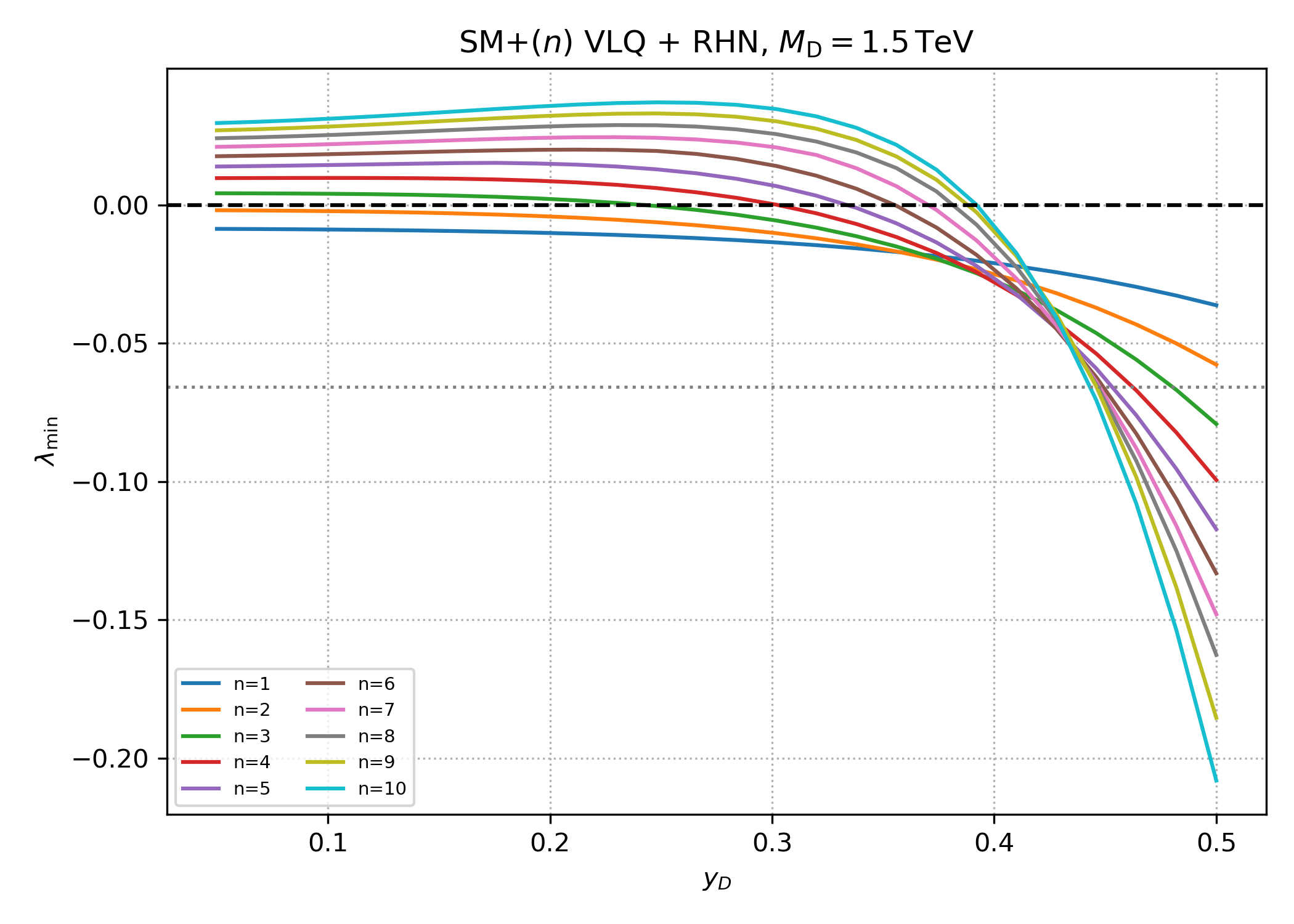}
        \\[2pt]
                {\small (a)}
    \end{minipage}\hfill
    %--- panel (b): MD = 3.0 TeV ---%
    \begin{minipage}[t]{0.5\textwidth}
        \centering
        \includegraphics[width=\linewidth]{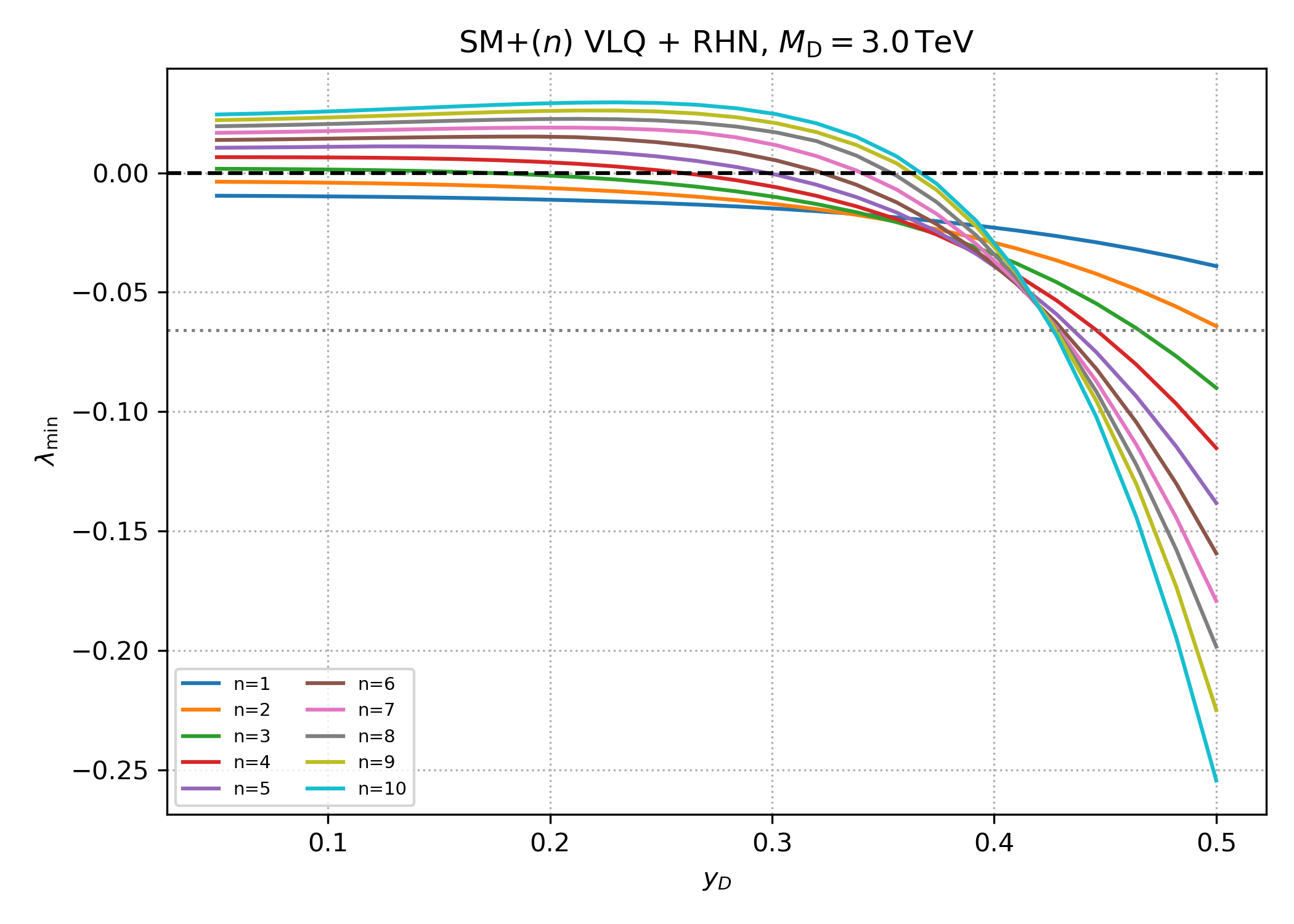}
        \\[2pt]
                {\small (b)}
    \end{minipage}\hfill
    %--- panel (c): MD = 5.0 TeV ---%
    \begin{minipage}[t]{0.5\textwidth}
        \centering
        \includegraphics[width=\linewidth]{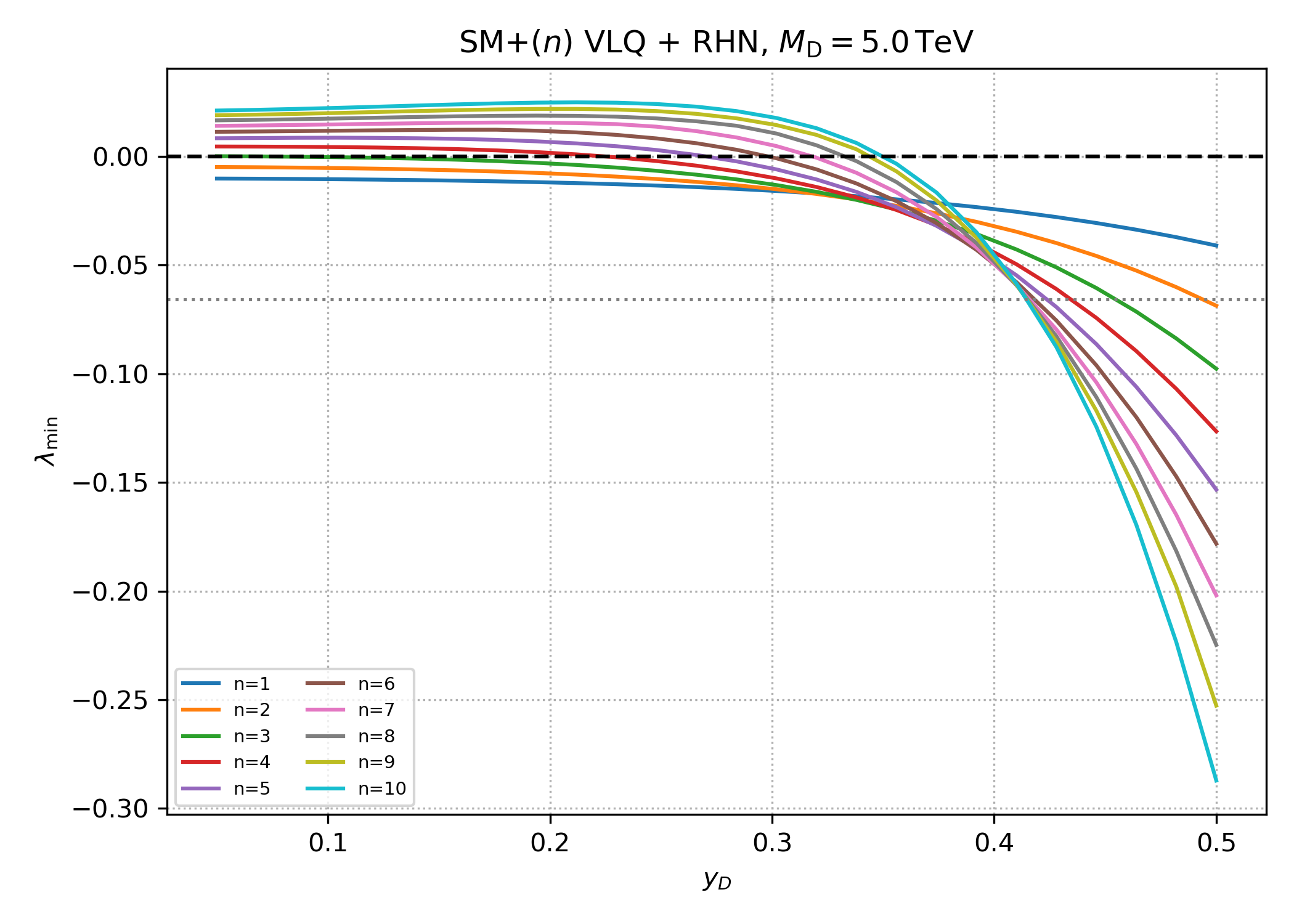}
        \\[2pt]
        {\small (c)}
    \end{minipage}

    \caption{
        Vacuum stability in the  SM+$(n)$ VLQ+RHN framework.
        Panels (a)–(c) show the minimum value $\lambda_{\min}$ as a function of
        the VLQ Yukawa coupling $y_D$ for benchmark masses
        $M_{\mathcal{D}} = 1.5,\,3.0,\,5.0$~TeV, respectively.
        In each panel, curves correspond to $n=1\text{--}10$.
    }
    \label{fig:vacuum_fullmodel}
\end{figure*}
\FloatBarrier

  \item At all three benchmark masses, the dependence on the number of VLQs $n$ shows 
a characteristic two–regime behaviour.  
For sufficiently small Yukawa couplings, increasing $n$ shifts $\lambda_{\min}$ 
upward and improves vacuum stability.  
However, beyond a certain $y_{\mathcal{D}}$ value in each panel, the destabilizing 
$\sim -\,n\,y_{\mathcal{D}}^4$ contribution dominates the running, and the trend reverses:  
larger $n$ then drives $\lambda_{\min}$ downward more rapidly as $y_{\mathcal{D}}$ 
increases.  
This turnover is clearly visible in all three plots and marks the point beyond 
which the number of VLQs $n$ no longer stabilizes the vacuum but instead 
accelerates its loss.

  \item In particular, the case $n=1$ and $n=2$ never yields a stable electroweak vacuum 
for any of the benchmark masses. For $n \ge 3$, vacuum stability can be achieved within a finite interval of 
  small $y_{\mathcal{D}}$, but this interval becomes progressively narrower as $M_{\mathcal{D}}$ is 
  increased.
\end{enumerate}

Taken together, these results show that, in the presence of a single RHN with 
$y_N = 0.42$, restoring  vacuum stability requires sufficiently 
small VLQ Yukawa couplings and, in practice, at least three or more down-type 
VLQs, with the case $n=3$ being viable only for the  benchmark masses 
$M_{\mathcal{D}} = 1.5, 3.0~\text{TeV}$.

Furthermore, it is well established that electroweak vacuum stability is highly sensitive to the top quark mass ($m_t$) due to its dominant role in the renormalization-group evolution of the Higgs quartic coupling. To assess the reliability of our stability predictions against experimental uncertainties in $m_t$, we evaluate $\lambda_{min}$ as a function of the top quark mass across its current PDG range ($172.56 \pm 0.31$ GeV)\cite{PDG2024}.

\begin{figure*}[h]
    \centering

    %--- panel (a): MD = 1.5 TeV ---%
    \begin{minipage}[t]{0.5\textwidth}
        \centering
        \includegraphics[width=\linewidth]{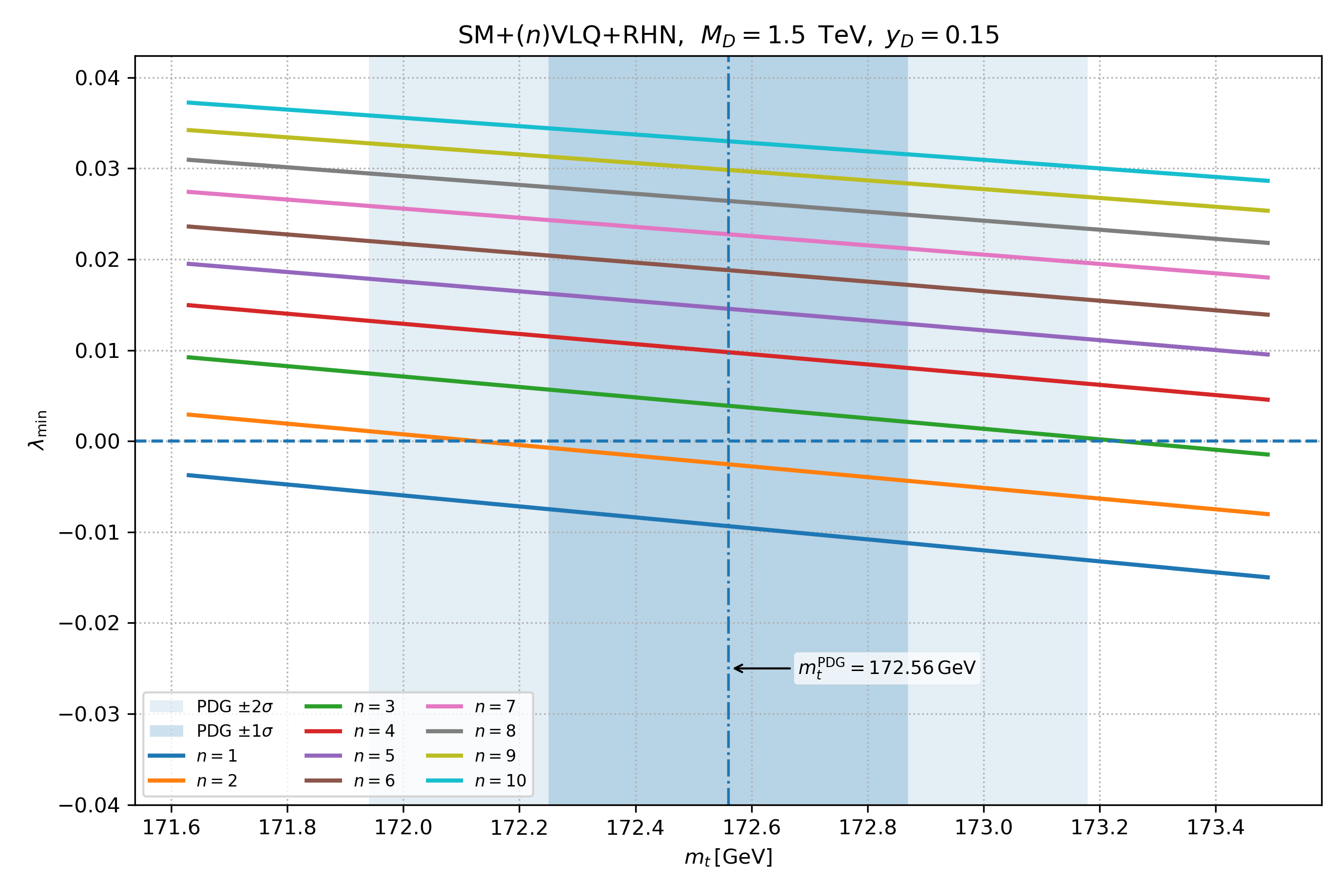}
        \\[2pt]
   
    \end{minipage}\hfill
    %--- panel (b): MD = 3.0 TeV ---%
    \begin{minipage}[t]{0.5\textwidth}
        \centering
        \includegraphics[width=\linewidth]{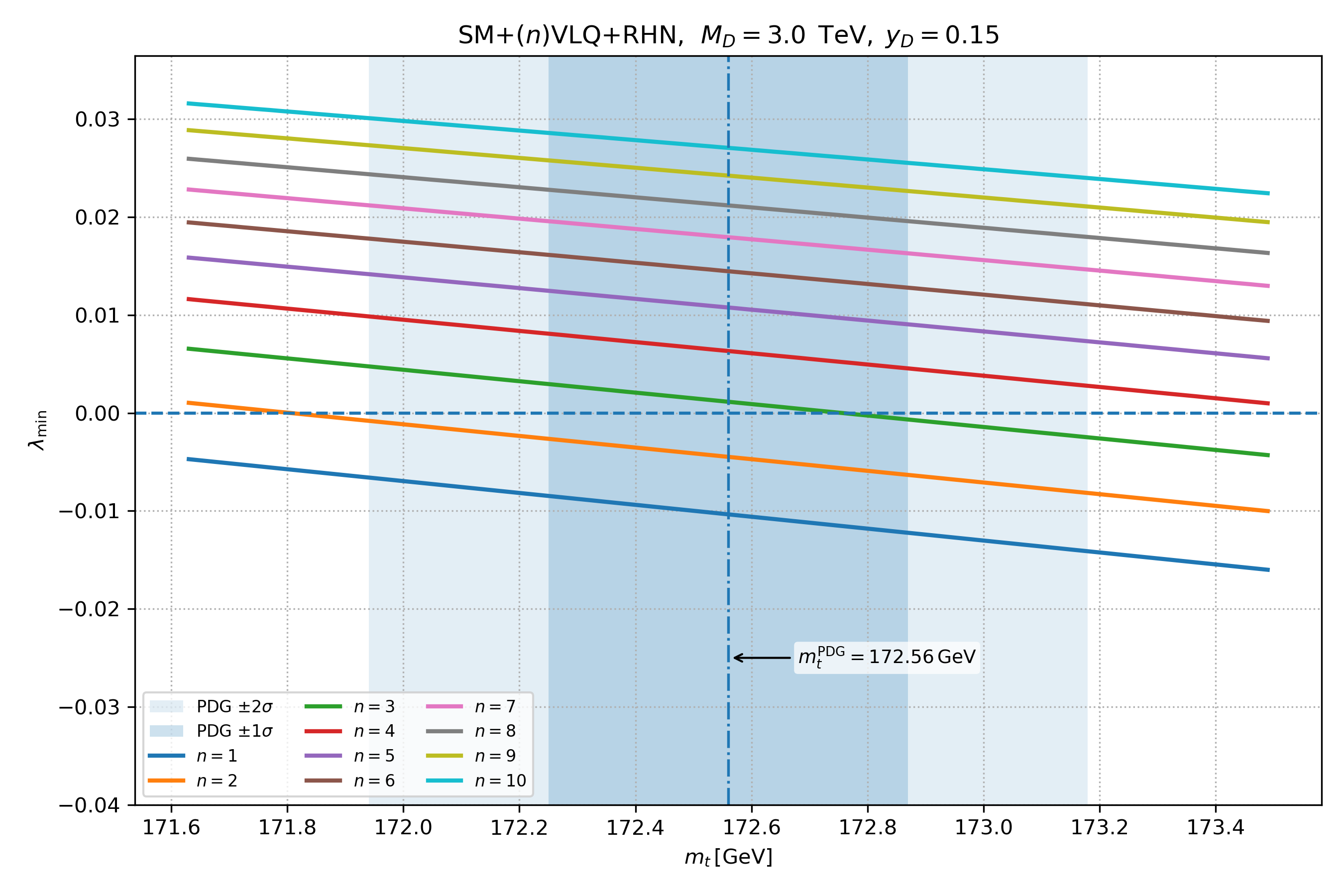}
        \\[2pt]
          
    \end{minipage}\hfill
    %--- panel (c): MD = 5.0 TeV ---%
    \begin{minipage}[t]{0.5\textwidth}
        \centering
        \includegraphics[width=\linewidth]{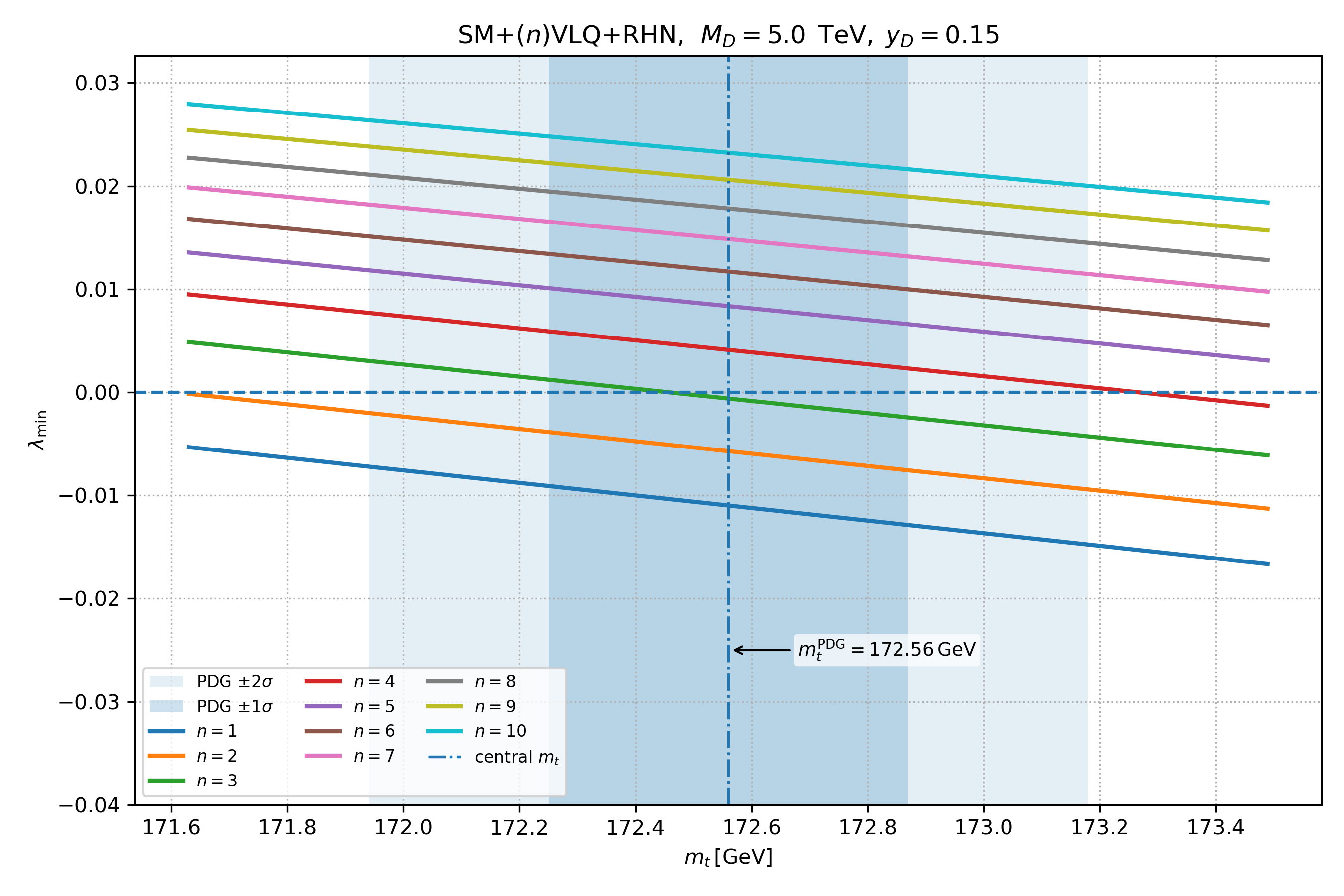}
        \\[2pt]
    
    \end{minipage}

    \caption{
        Dependence of the minimum value of the Higgs quartic coupling, $\lambda_{\min}$, 
on the top quark mass $m_t$ in the SM+$ (n)$VLQ+RHN framework with 
$M_\mathcal{D}=1.5,; 3.0,;5.0 ~\mathrm{TeV}$, $y_D=0.15$, and $y_N=0.42$. 
The vertical dashed line corresponds to the PDG average value of the top quark mass, 
while the shaded bands indicate the $1\sigma$ and $2\sigma$ uncertainty ranges. 
Each curve represents a different number of vector-like quarks $n=1$--$10$. 
    }
    \label{fig:mt_1}
\end{figure*}

In Fig.~\ref{fig:mt_1}, $\lambda_{\min}$ is shown as a function of $m_t$ for different numbers of VLQs ($n=1$–$10$), for the benchmark masses $M_{\mathcal{D}} = 1.5,; 3.0,; 5.0~\mathrm{TeV}$, with $y_{\mathcal{D}}=0.15$ and $y_N=0.42$ fixed. The dark and light shaded blue regions indicate the $1\sigma$ and $2\sigma$ experimental intervals for $m_t$, respectively. As expected, $\lambda_{\min}$ exhibits a near-linear decrease with increasing $m_t$. Within the experimentally allowed $m_t$ range, the stability status of most configurations remains unchanged. However, specific cases are highly sensitive to the exact $m_t$ value: the $n=2$ configuration for $M_{\mathcal{D}} = 1.5~\mathrm{TeV}$, and the $n=3$ configurations for $M_{\mathcal{D}} = 3.0$ and $5.0~\mathrm{TeV}$, cross the $\lambda_{\min}=0$ threshold, transitioning between stable and metastable regimes depending on the $1\sigma$ and $2\sigma$ limits. Crucially, for $n \ge 4$, absolute vacuum stability is robustly maintained across the entire $2\sigma$ interval for all considered VLQ masses. This confirms that the inclusion of at least four VLQs provides a reliable stabilization mechanism within the considered parameter space, effectively independent of current top-quark mass uncertainties.

\FloatBarrier

\begin{figure*}[h]
    \centering

    %--- panel (a): M_D = 1.5 TeV ---%
    \begin{minipage}[t]{0.48\textwidth}
        \centering
        \includegraphics[width=\linewidth]{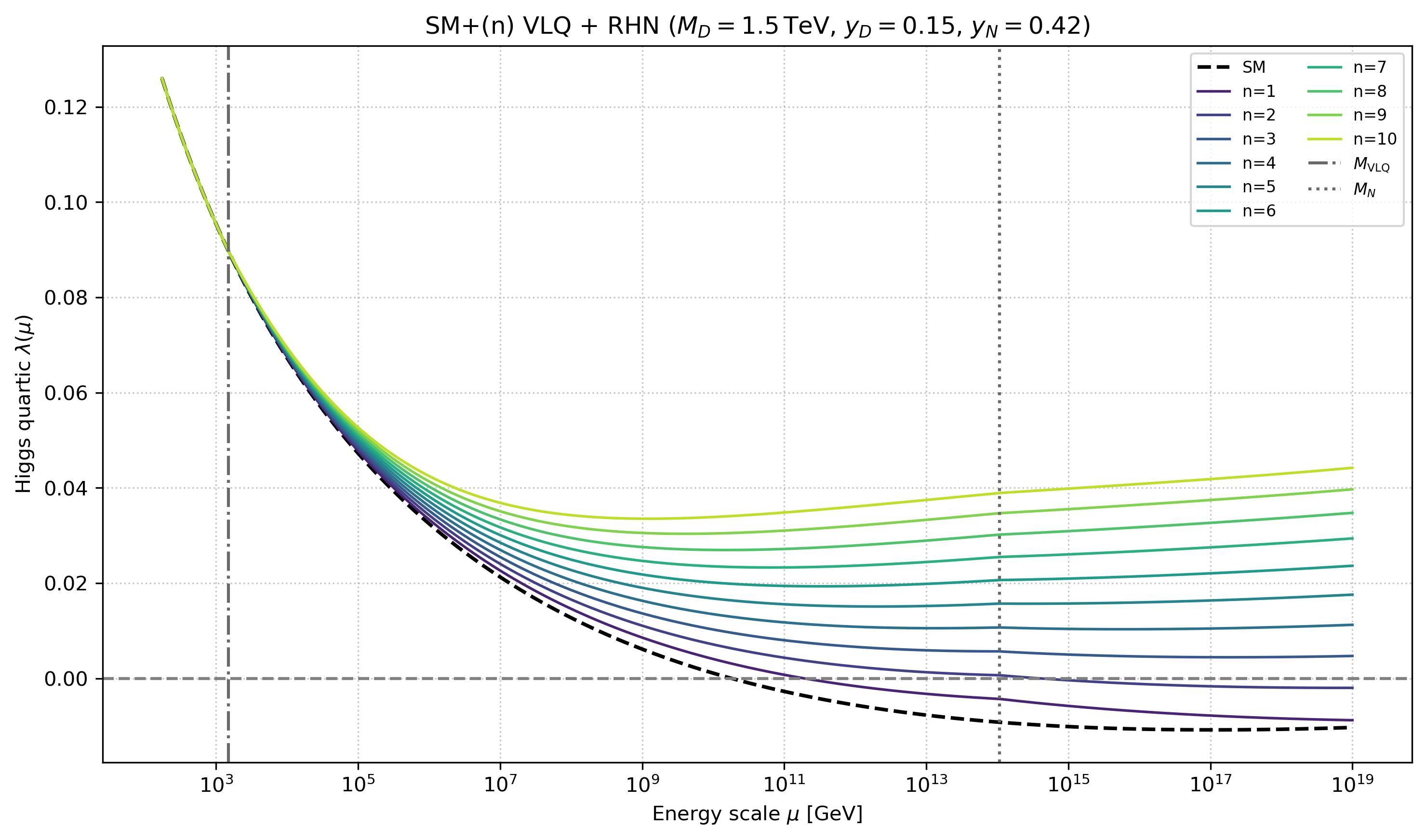}
        \\[2pt]
        {\small (a)}
    \end{minipage}
    \hfill
    %--- panel (b): M_D = 3.0 TeV ---%
    \begin{minipage}[t]{0.48\textwidth}
        \centering
        \includegraphics[width=\linewidth]{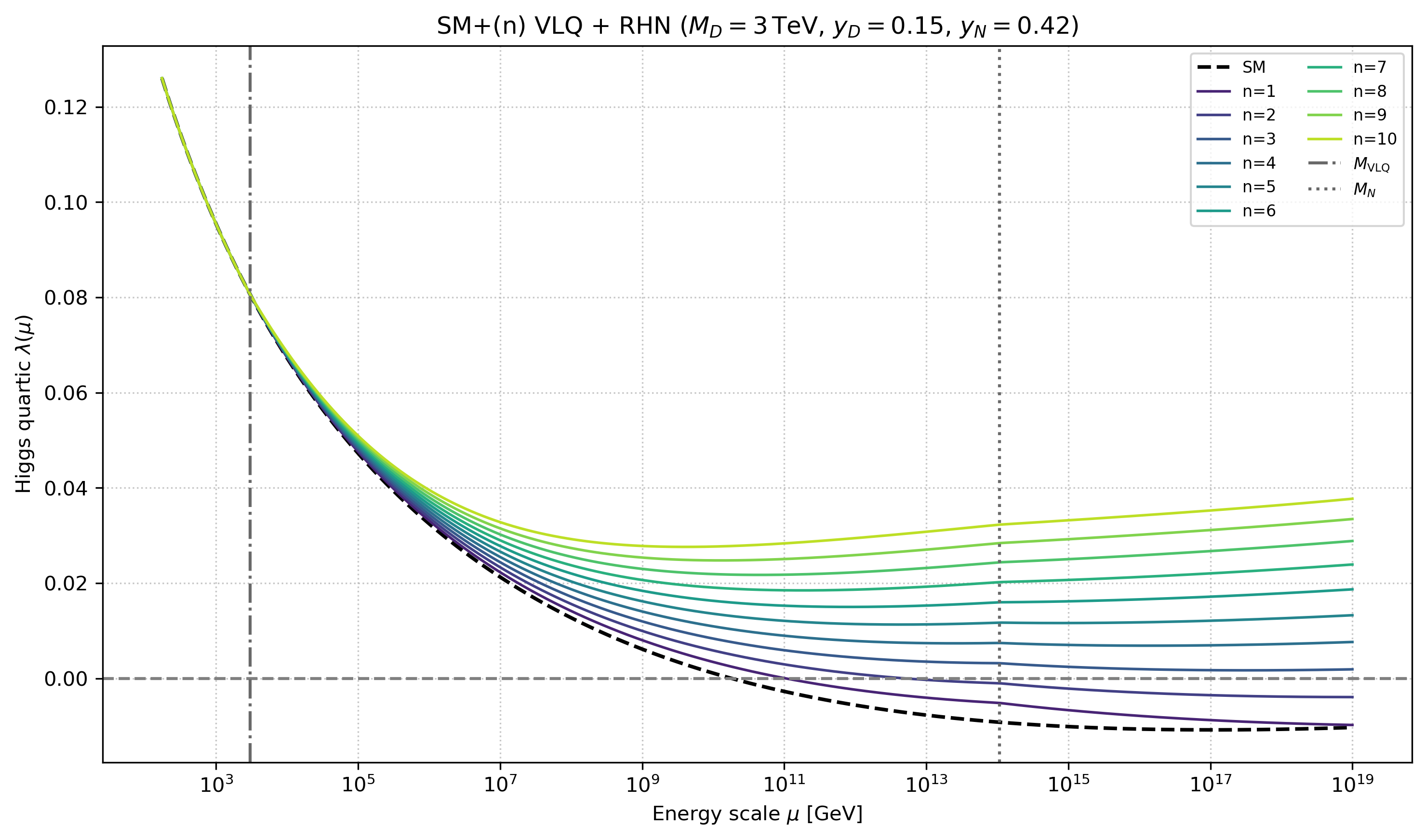}
        \\[2pt]
        {\small (b)}
    \end{minipage}

    \vspace{8pt}

    %--- panel (c): M_D = 5.0 TeV ---%
    \begin{minipage}[t]{0.48\textwidth}
        \centering
        \includegraphics[width=\linewidth]{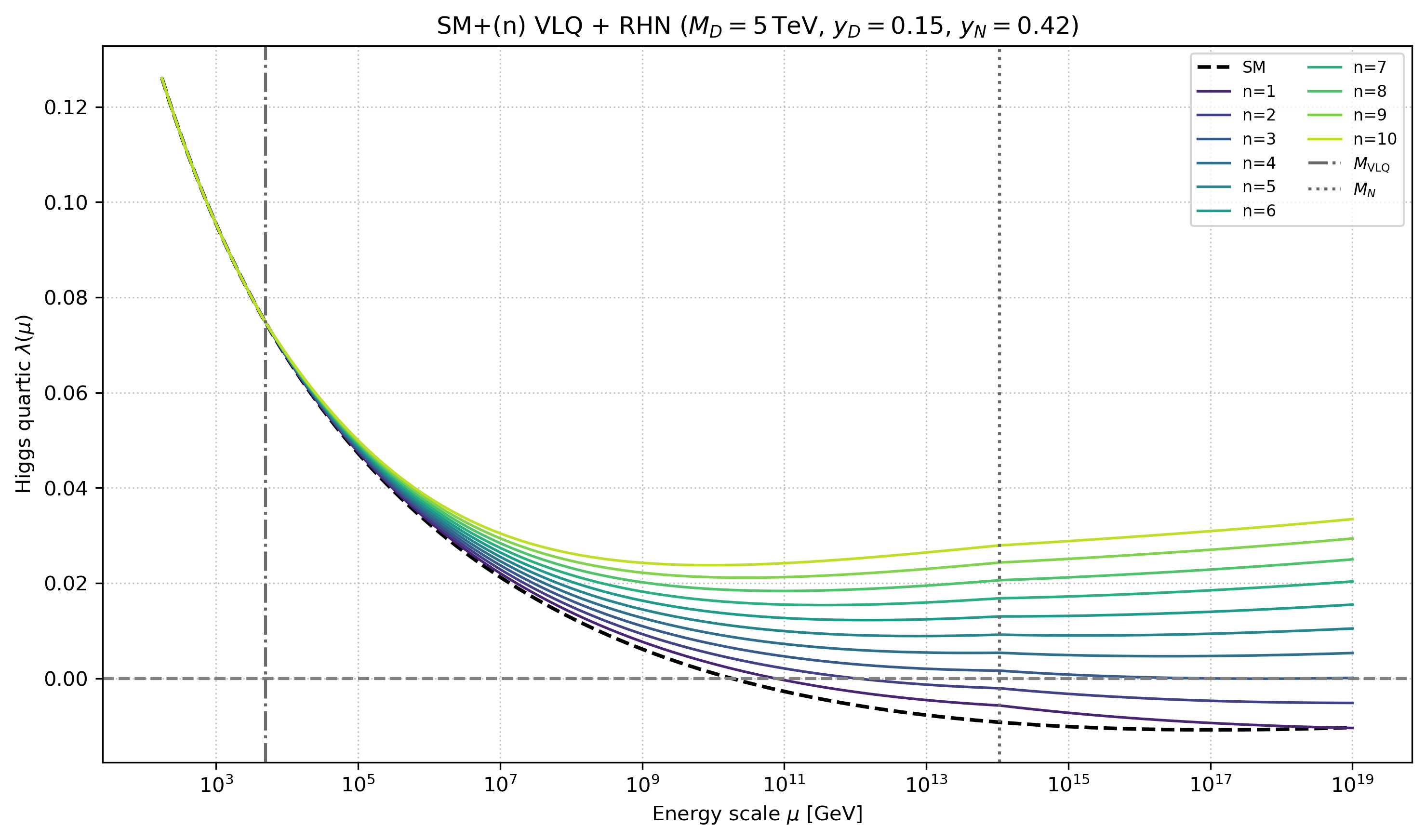}
        \\[2pt]
        {\small (c) }
    \end{minipage}

    \caption{Running of the Higgs quartic coupling $\lambda(\mu)$ in the SM+$(n)$VLQ+RHN model for the benchmark VLQ masses
$M_{\mathcal{D}} = 1.5,;3.0,;5.0~\mathrm{TeV}$.
    The curves correspond to $n=1$--$10$ for fixed $y_{\mathcal{D}} = 0.15$ and 
    $y_N = 0.42$, while the dashed black curve shows the SM result. 
    The vertical dashed and dotted lines indicate the VLQ and RHN mass
    thresholds, respectively.}
    \label{fig:lambda_running_SM_nVLQ_RHN}
\end{figure*}

To illustrate the scale evolution of the Higgs quartic coupling more 
transparently, we select a common Yukawa value ($y_{\mathcal{D}=0.15}$)for all benchmark masses and the world average value of the top quark mass as reported by the PDG ($m_t^{PDG}=172.56$)  ~\cite{PDG2024}. 
Since vacuum stabilization with the smallest possible number of VLQs
is achieved only in the region of relatively small VLQ Yukawa couplings,
we adopt a conservative choice $y_{\mathcal{D}} = 0.15$.
This value lies within the experimentally allowed region for all benchmark
masses according to Eq.(\ref{theta}) and ensures that the stabilizing effect of the VLQ sector can already
be realized for the minimal viable $n$, enabling a meaningful
comparison of the running of $\lambda(\mu)$ as a function of the number of VLQs. The running of the Higgs quartic coupling as a function of the renormalization scale is presented in Fig.(\ref{fig:lambda_running_SM_nVLQ_RHN}) for the benchmark VLQ mass values $M_{\mathcal{D}} = 1.5,\, 3.0,\, 5.0$ and $5~\mathrm{TeV}$. \\
As  indicated by the stability limits in Fig.\ref{fig:vacuum_fullmodel}, the case $n=1$ and $n=2$
fail to stabilize the Higgs potential for any of the benchmark masses considered, 
while $n=3$ achieves stability only for $M_D = 1.5$ and $3.0~\text{TeV}$. 
In contrast, for $n \ge 4$ all three benchmark mass values successfully maintain 
a positive Higgs quartic coupling up to the Planck scale.
Fig.~(\ref{fig:lambda_running_SM_nVLQ_RHN}) further illustrates this behavior:
as the number of VLQs $n$ increases, their contribution to the
RG evolution shifts the running of $\lambda(\mu)$ toward
larger values, such that the Higgs quartic coupling remains positive up to the Planck scale.

\FloatBarrier

Moreover, the evolution of $\lambda(\mu)$ in the inflationary domain 
$10^{15}\!-\!10^{19}~\text{GeV}$ remains remarkably smooth, 
with no indication of rapid variations or destabilizing trends. 
This controlled high-scale behaviour reflects a well--balanced interplay between 
the destabilizing RHN Yukawa contribution and the stabilizing effect of the VLQ sector. 
Such mild running of the quartic coupling is precisely the regime required for 
a viable Higgs--inflation setup, where the flatness and stability of the 
effective potential are sensitive to the ultraviolet behaviour of $\lambda(\mu)$.

\subsection{Vacuum Stability in the $SM+(n)VLQ$ Framework } 

In the absence of RHN, the stabilizing effect of the VLQs is significantly 
stronger: the allowed region in $y_D$ is wider and the values of $\lambda_{\min}$ 
are consistently higher.  
This comparison highlights the non-trivial impact of individual fermionic degrees of freedom on the high-scale behavior of the effective potential.

Following the same procedure as in the SM+$(n)$VLQ+RHN framework discussed in the previous section, we analyze the dependence of Higgs vacuum stability on the VLQ Yukawa coupling $y_D$.
For the benchmark masses $M_D = 1.5,\;3.0,\;5.0~\mathrm{TeV}$, the RGEs are evolved up to the Planck scale, and the corresponding minimum value of the Higgs quartic coupling, $\lambda_{\min}$, is extracted.

Fig.\ref{fig:vacuum_limitmodel} displays the minimum value of the Higgs quartic coupling, $\lambda_{\min}$, as a function of the VLQ Yukawa coupling $y_D$, for $n = 1\text{--}10$ and for the benchmark mass choices $M_D = 1.5,;3.0,;5.0~\mathrm{TeV}$.
Compared with the SM+$(n)$VLQ+RHN case discussed previously, the qualitative 
behaviour remains similar, but important differences appear because the 
destabilizing RHN Yukawa contribution is now absent. The main features of Fig.\ref{fig:vacuum_limitmodel} can be summarized as follows:
\begin{enumerate}
\item
Removing the RHN Yukawa eliminates the negative contribution that previously 
pushed  $\lambda_{\min}$ downward at high scales.  
Consequently, for each $n$,  $\lambda_{\min}$ stays higher than in the full 
SM+$(n)$VLQ+RHN model.  
This confirms that VLQs alone have a net stabilizing tendency at small 
Yukawa couplings. 

\item
As in the full model, increasing the number of VLQs shifts the running 
upward, delaying (or preventing) the point at which  $\lambda_{\min}$ turns 
negative.  
However, without the RHN, this stabilizing effect is  stronger and 
persists to larger values of $y_\mathcal{D}$.

\begin{figure*}[h]
    \centering

    %--- panel (a): MD = 1.5 TeV ---%
    \begin{minipage}[t]{0.5\textwidth}
        \centering
        \includegraphics[width=\linewidth]{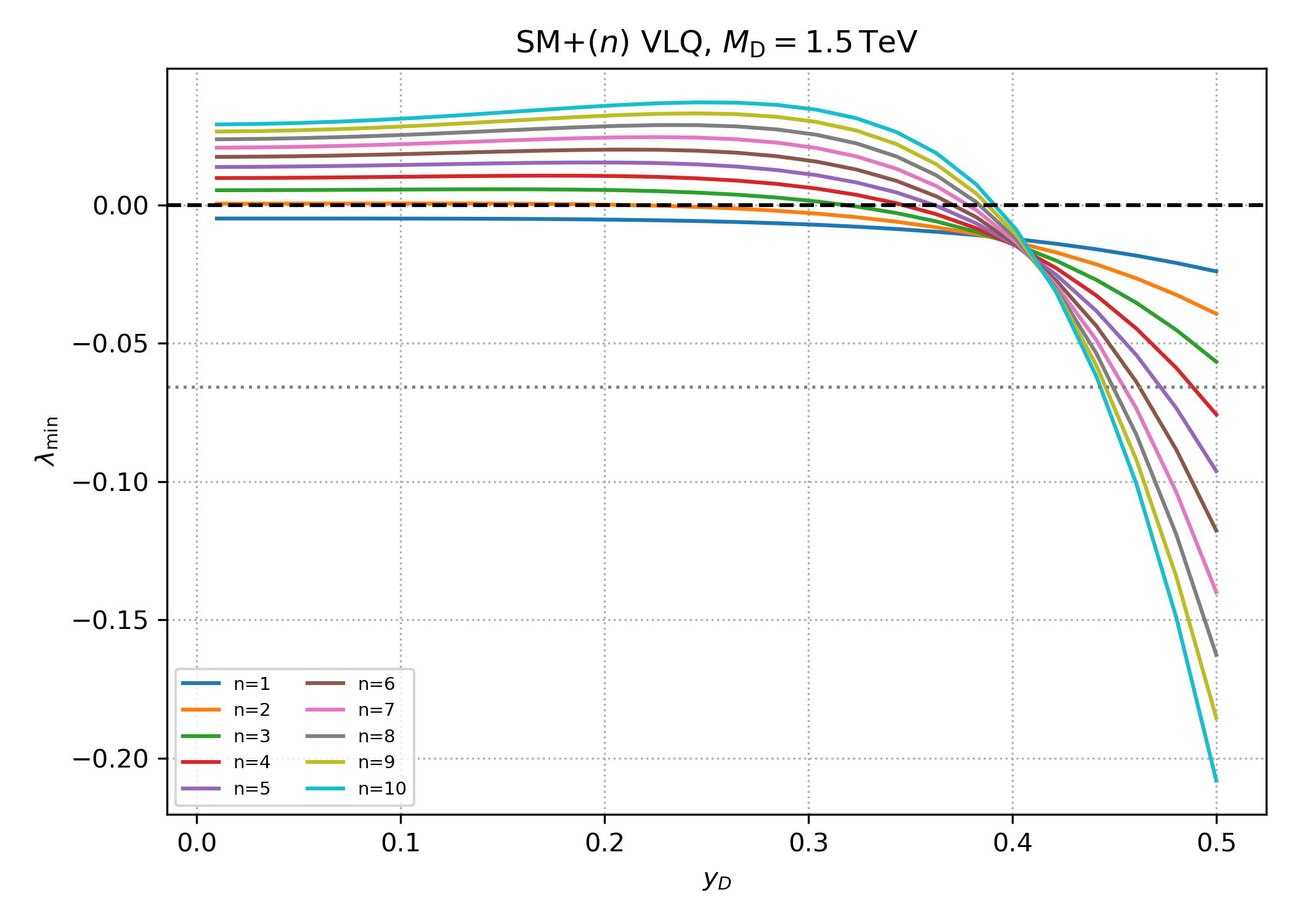}
        \\[2pt]
        {\small (a)}
    \end{minipage}\hfill
    %--- panel (b): MD = 3.0 TeV ---%
    \begin{minipage}[t]{0.5\textwidth}
        \centering
        \includegraphics[width=\linewidth]{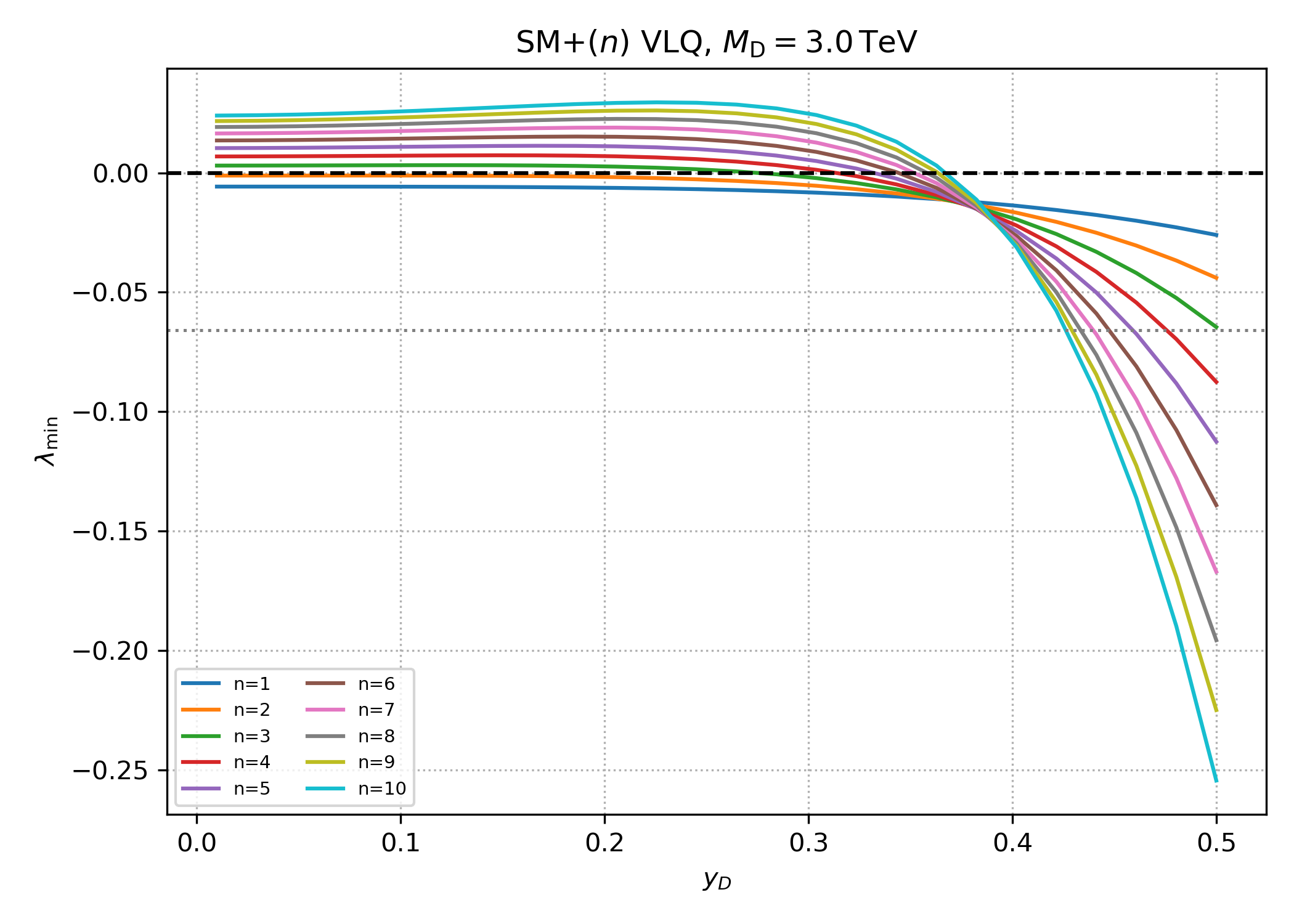}
        \\[2pt]
        {\small (b)}
    \end{minipage}\hfill
    %--- panel (c): MD = 5.0 TeV ---%
    \begin{minipage}[t]{0.5\textwidth}
        \centering
        \includegraphics[width=\linewidth]{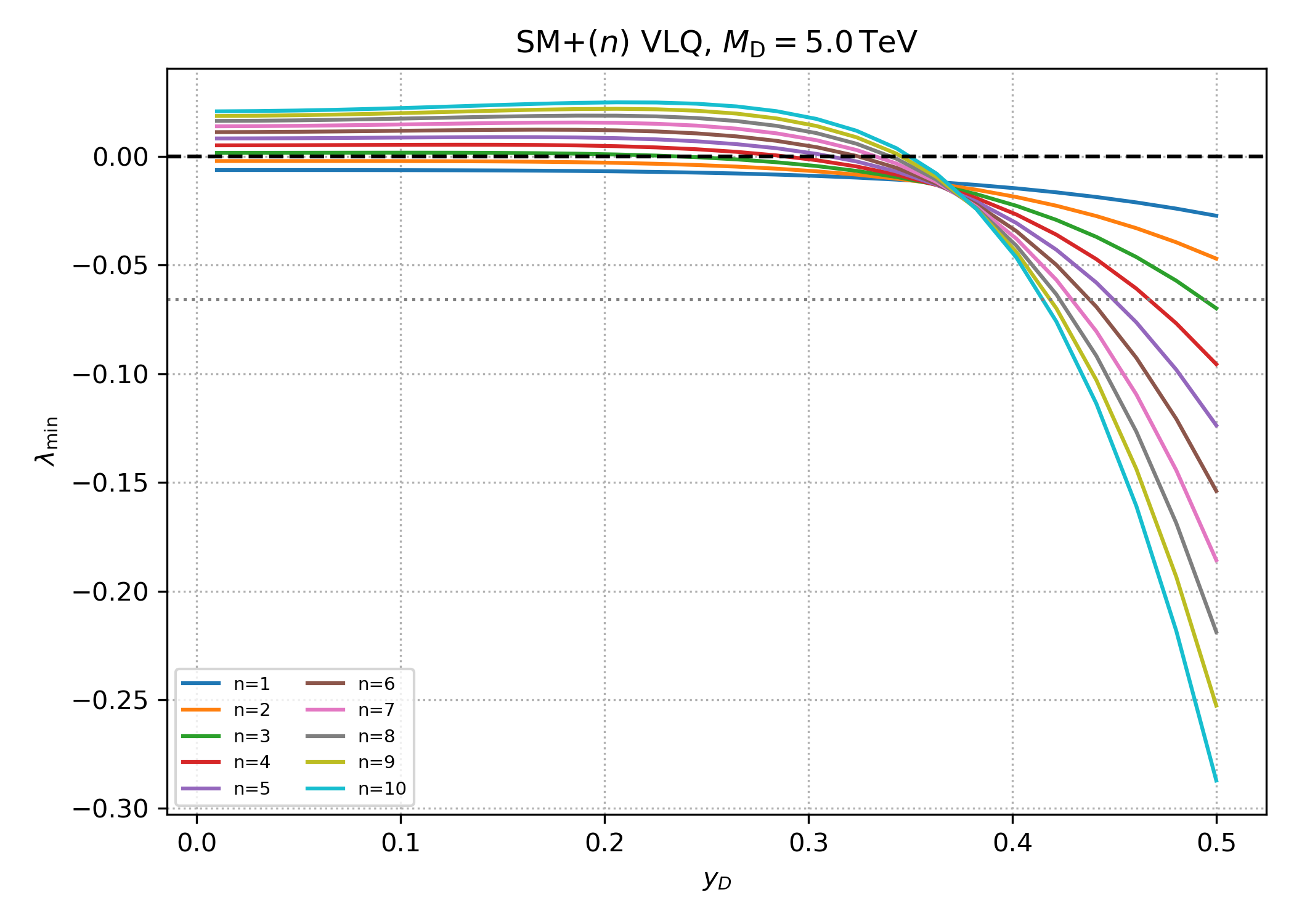}
        \\[2pt]
        {\small (c)}
    \end{minipage}

    \caption{
        Vacuum stability in the SM+$(n)$ VLQ framework.
        Panels (a)–(c) show the minimum value $\lambda_{\min}$ as a function of
        the VLQ Yukawa coupling $y_D$ for benchmark masses
        $M_{\mathcal{D}} = 1.5,\,3.0,\,5.0$~TeV, respectively.
        In each panel, curves correspond to $n=1\text{--}10$.
    }
    \label{fig:vacuum_limitmodel}
\end{figure*}
\FloatBarrier

\item
The characteristic transition seen previously --- where large values of 
$n$ eventually accelerate the fall of  $\lambda_{\min}$ still appears, but with a reduced rate and at later scales.  
In other words, the destabilizing regime sets in more softly because only 
the VLQ sector drives it, not the RHN.

\item
In the SM+$(n)$VLQ+RHN case, $n=1$ and $n=2$  never achieved stability 
in the allowed parameter space.  
Here, for SM+$(n)$VLQ, even small $n$ values lead to a visibly less negative 
$\lambda_{\min}$, and several trajectories remain close to or above zero. For $n=2$, a vacuum--stable region exists only for the lightest VLQ mass 
  $M_{\mathcal{D}} = 1.5~\text{TeV}$, while for $M_{\mathcal{D}} = 3.0$ and $5.0~\text{TeV}$ the 
  $n=2$ curve stays in the metastable regime.  
Models with $n \ge 3$ are substantially more stable than in the full model, 
confirming that the RHN Yukawa was the primary destabilizing agent.

\item
Overall, the SM+$({n})$VLQ scenario leads to improved electroweak vacuum stability compared to the SM+$(n)$VLQ+RHN model for all considered values of $n$ and $M_{\mathcal{D}}$, as reflected by larger values of $\lambda_{\min}$ and a broader region with $\lambda_{\min}>0$.
In this case, vacuum stability is still governed primarily by the interplay between
$n$ and $y_\mathcal{D}$.
The absence of the RHN Yukawa contribution generally enlarges the stability region,
particularly for larger values of $n$.

\end{enumerate}

In parallel with the analysis of the SM+$(n)$VLQ+RHN framework, we also evaluate the resilience of our stability predictions against the experimental uncertainties of $m_t$ in the absence of the RHN. To this end, we perform a similar scan of $\lambda_{\min}$ as a function of the top quark mass across its current PDG range ($172.56 \pm 0.31$ GeV) \cite{PDG2024}.

\begin{figure*}[h!]
    \centering

    %--- panel (a): MD = 1.5 TeV ---%
    \begin{minipage}[t]{0.46\textwidth}
        \centering
        \includegraphics[width=\linewidth]{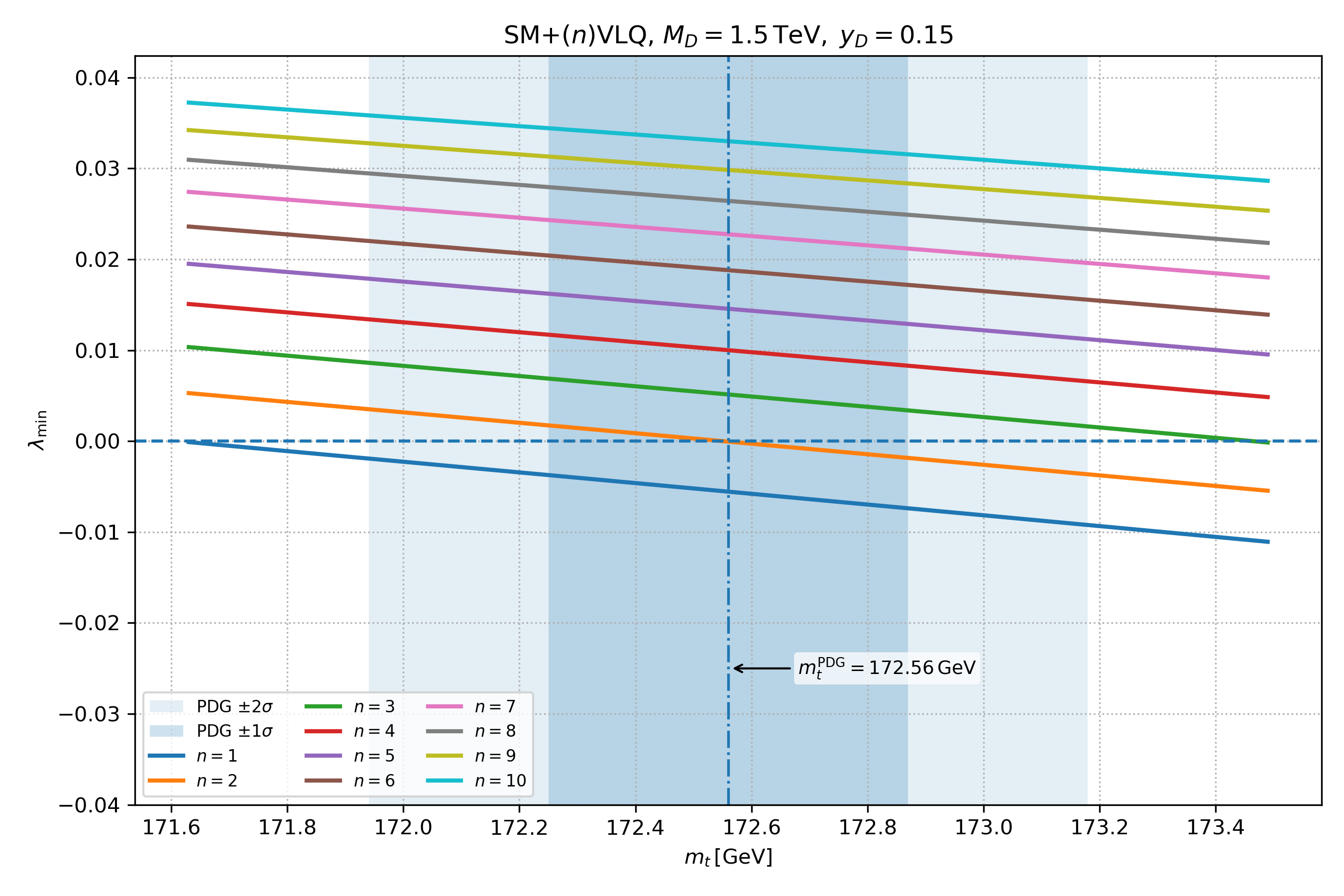}
        \\[2pt]
   
    \end{minipage}\hfill
    %--- panel (b): MD = 3.0 TeV ---%
    \begin{minipage}[t]{0.46\textwidth}
        \centering
        \includegraphics[width=\linewidth]{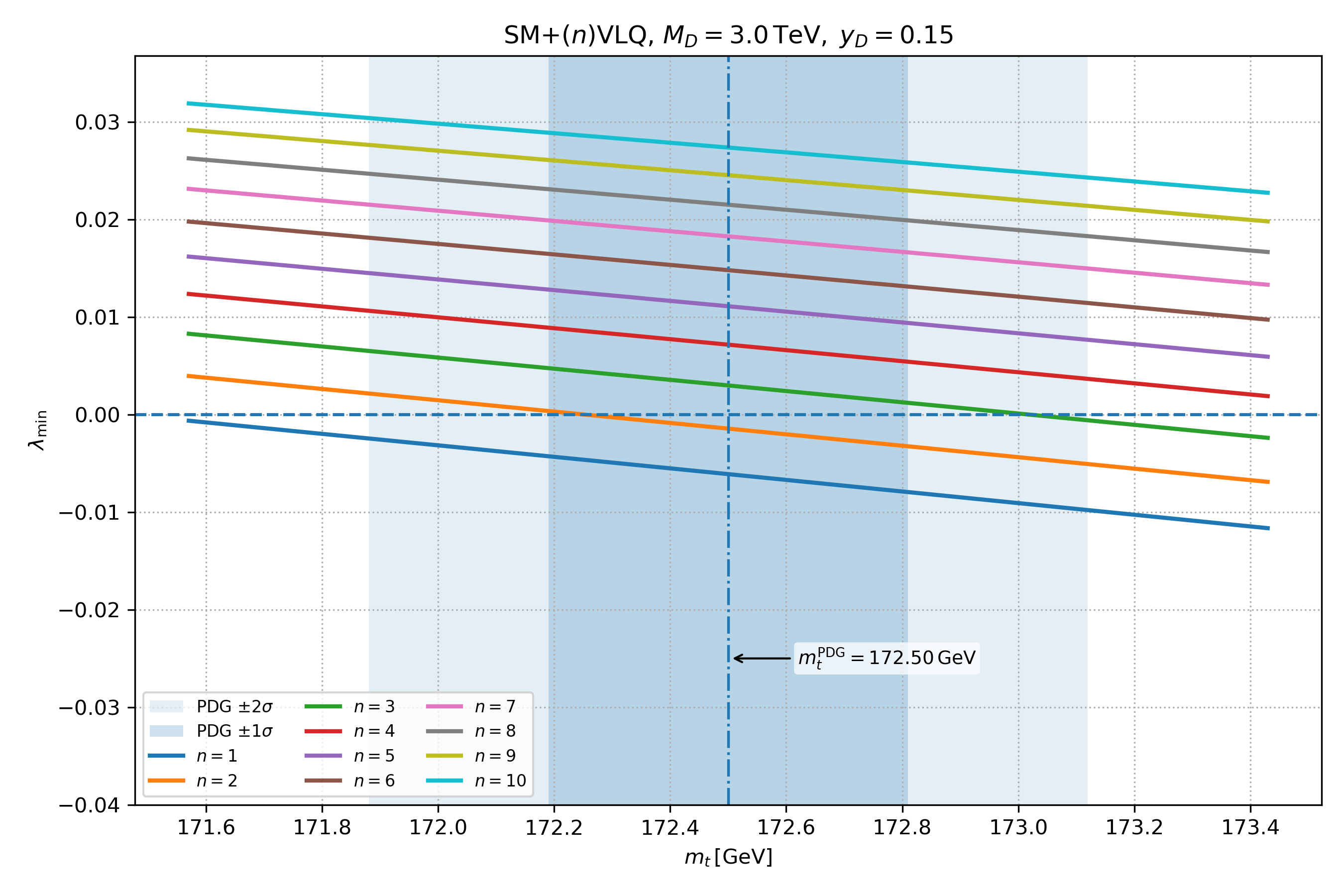}
        \\[2pt]
          
    \end{minipage}\hfill
    %--- panel (c): MD = 5.0 TeV ---%
    \begin{minipage}[t]{0.46\textwidth}
        \centering
        \includegraphics[width=\linewidth]{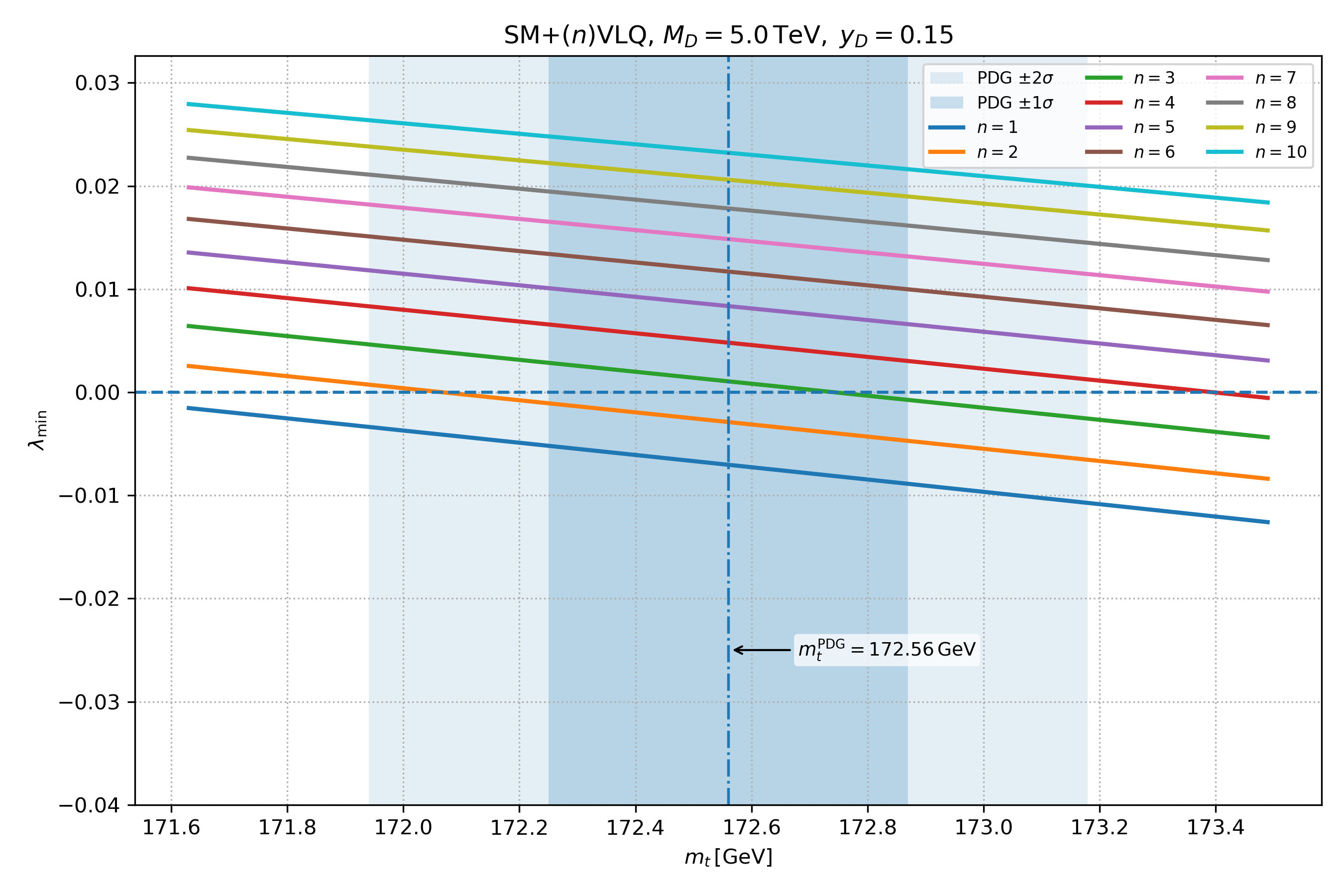}
        \\[2pt]
    
    \end{minipage}

    \caption{
        Dependence of the minimum value of the Higgs quartic coupling, $\lambda_{\min}$, 
on the top quark mass $m_t$ in the SM+$ (n)$VLQ framework with 
$M_\mathcal{D}=1.5,; 3.0,;5.0 ~\mathrm{TeV}$, $y_D=0.15$, and $y_N=0.42$. 
The vertical dashed line corresponds to the PDG average value of the top quark mass, 
while the shaded bands indicate the $1\sigma$ and $2\sigma$ uncertainty ranges. 
Each curve represents a different number of vector-like quarks $n=1$--$10$. 
    }
    \label{fig:mt_2}
\end{figure*}

In Fig.~\ref{fig:mt_2}, we illustrate this dependence within the SM+($n$)VLQ framework for the benchmark masses $M_{\mathcal{D}} = 1.5$, $3.0$, and $5.0~\mathrm{TeV}$, with $y_{\mathcal{D}}=0.15$. As dictated by the renormalization-group equations, larger values of $m_t$ increase the negative contribution of the top Yukawa coupling, leading to a near-linear decrease in $\lambda_{\min}$. This $m_t$ dependence plays a critical role in determining the exact number of VLQs required to achieve an absolutely stable vacuum. Specifically, within the $1\sigma$ and $2\sigma$ experimental intervals of $m_t$, the stability boundaries are highly sensitive to the top quark mass: the $n=2$ configuration for $M_{\mathcal{D}} = 1.5~\mathrm{TeV}$, and the $n=3$ configurations for $M_{\mathcal{D}} = 3.0$ and $5.0~\mathrm{TeV}$, cross the $\lambda_{\min}=0$ threshold, transitioning from stable to metastable regimes as $m_t$ increases. In contrast, configurations with $n \ge 4$ safely maintain $\lambda_{\min} > 0$ across the entire experimentally allowed $m_t$ range for all evaluated VLQ masses. This analysis demonstrates that while the absolute stability boundary for smaller VLQ multiplicities is sensitive to current $m_t$ uncertainties, the inclusion of at least four VLQs ensures a robust stabilization independent of these variations.

\FloatBarrier

\begin{figure*}[h]
    \centering

    %--- panel (a): M_D = 1.5 TeV ---%
    \begin{minipage}[t]{0.48\textwidth}
        \centering
        \includegraphics[width=\linewidth]{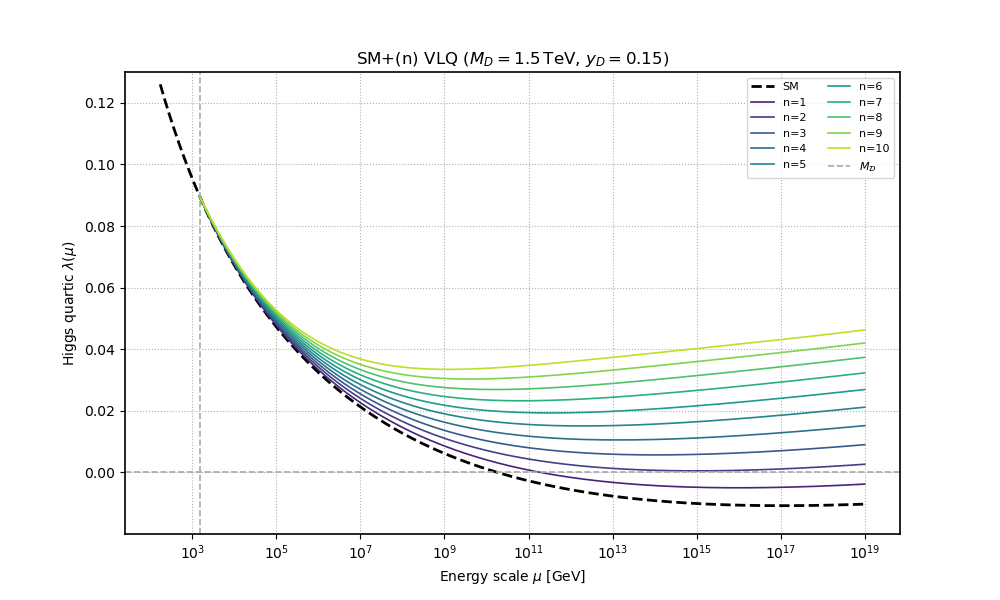}
        \\[2pt]
                {\small (a)}
    \end{minipage}
    \hfill
    %--- panel (b): M_D = 3.0 TeV ---%
    \begin{minipage}[t]{0.48\textwidth}
        \centering
        \includegraphics[width=\linewidth]{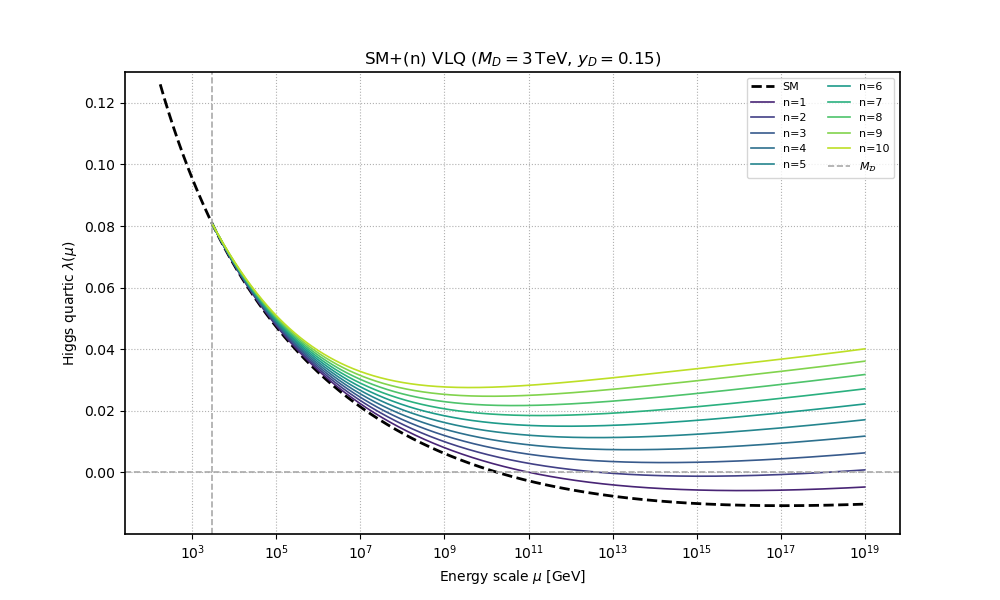}
        \\[2pt]
        {\small (b)}
    \end{minipage}

    \vspace{8pt}

    %--- panel (c): M_D = 5.0 TeV ---%
    \begin{minipage}[t]{0.48\textwidth}
        \centering
        \includegraphics[width=\linewidth]{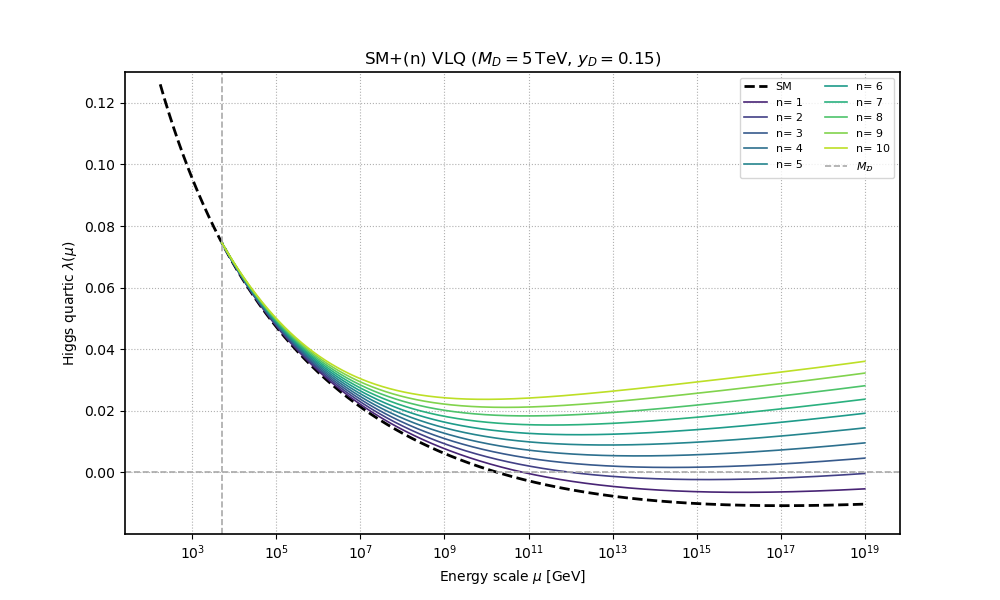}
        \\[2pt]
                {\small (c)}
    \end{minipage}

    \caption{
    Running of the Higgs quartic coupling $\lambda(\mu)$ in the SM+$(n)$VLQ framework for the benchmark VLQ masses
$M_{\mathcal{D}} = 1.5,;3.0,;5.0~\mathrm{TeV}$.
    The curves correspond to $n=1$--$10$ for fixed $y_{\mathcal{D}} = 0.15$, while the dashed black curve shows the SM result. 
    The vertical dashed line indicates the VLQ  mass
    threshold.
    }
    \label{fig:lambda_running_three_panels}
\end{figure*}

To illustrate the scale dependence of the Higgs quartic coupling in a transparent manner, we fix the VLQ Yukawa coupling to a common benchmark value, $y_\mathcal{D}= 0.15$, for all considered VLQ masses.
The resulting running of the Higgs quartic coupling, $\lambda(\mu)$, is shown in Fig.\ref{fig:lambda_running_three_panels} for the benchmark masses
$M_D = 1.5,\;3.0,\;5.0~\mathrm{TeV}$.

\FloatBarrier

As seen from the Fig.\ref{fig:lambda_running_three_panels}, the case $n=1$ is insufficient to stabilize the Higgs potential for any of the benchmark masses considered, while $n=2$ leads to vacuum stability only for the lower mass values $M_D = 1.5~\mathrm{TeV}$.
For $n \geq 3$, all benchmark masses yield a positive Higgs quartic coupling up to the Planck scale.

In comparison with the full SM+$(n)$VLQ+RHN framework, the SM+$(n)$VLQ model 
exhibits a noticeably different high--scale behaviour.  
Although the addition of VLQs alone lifts the running of $\lambda(\mu)$ and 
delays the destabilization of the potential, the quartic coupling develops an
enhanced curvature at intermediate and high scales.  By contrast, the SM+$(n)$VLQ+RHN scenario displays a remarkably smooth 
evolution of $\lambda(\mu)$ throughout the inflationary domain 
$10^{15}$--$10^{19}~\text{GeV}$.  
This smoothness does not arise from the VLQ contribution alone; rather, it 
is a consequence of the delicate balance between the stabilizing effect of 
the VLQ sector and the destabilizing RHN Yukawa interaction.  
These competing contributions partially cancel, yielding a controlled and 
slowly varying high--scale running of the quartic coupling with no abrupt 
changes in slope.  
Such regulated ultraviolet behaviour is precisely the regime required for a 
viable Higgs--inflation setup, where the flatness and stability of the 
effective potential are highly sensitive to the evolution of $\lambda(\mu)$.

In the  SM+$(n)$VLQ+RHN framework, the absence of this balancing mechanism leads to a less regulated high--scale 
trajectory.  
While the vacuum--stability region is generally enlarged once the RHN 
contribution is removed, the smooth inflationary behaviour characteristic of 
the full SM+$({n})$VLQ+RHN model is no longer reproduced.

These vacuum stability results provide the foundation for the Higgs inflation 
analysis in Sec.~\ref{higgs-inf}, where the requirement of a positive 
RG-improved quartic coupling plays a central role in determining the 
inflationary potential.

\section{Higgs inflation}
\label{higgs-inf}

In our framework the radial mode of the Standard Model Higgs doublet plays the role of the inflaton at large field values. The Higgs field is non-minimally coupled to gravity in the metric formulation via
\begin{equation}
  S \supset \int d^4x \sqrt{-g} \left[
    - \frac{M_{\rm Pl}^2}{2}\, R
    - \xi\, H^\dagger H\, R
    + (D_\mu H)^\dagger (D^\mu H)
    - V(H,\mu)
  \right],
\end{equation}
where $M_{\rm Pl}$ is the reduced Planck mass, $\xi$ is the non-minimal Higgs--gravity coupling, and $V(H,\mu)$ denotes the renormalization-group (RG) improved Higgs potential. Working in unitary gauge and denoting the real Higgs field by $\phi$, we write the Jordan-frame potential as
\begin{equation}
  V_J(\phi,\mu) \;=\; \frac{1}{4}\,\lambda(\mu)\,\phi^4,
\label{eq:VJ_def}
\end{equation}
where $\lambda(\mu)$ is the running Higgs quartic coupling obtained from the full two-loop SM RGEs supplemented by the one-loop contributions of the vector-like quarks and the right-handed neutrino, with appropriate threshold matching at $\mu = M_{\mathcal{D}}$ and $\mu = M_N$. n the inflationary regime, we adopt the commonly used renormalization prescription in which the running scale is identified with the Higgs field value
\begin{equation}
  \mu \;\simeq\; \phi,
\end{equation}
This choice corresponds to the so-called Prescription II in the literature, where the renormalization scale is defined in the Jordan frame to reflect the field-dependent masses of particles propagating in loops\cite{Bezrukov:2009, Hamada:2014}.This prescription minimizes large logarithmic corrections of the form $\ln(\phi^2/\mu^2)$ in the RG-improved effective potential, thereby ensuring a consistent perturbative expansion and a stable inflationary trajectory. We restrict the analysis to the field range where $\lambda(\phi) > 0$ and $\phi$ lies between $10^{15}\,\text{GeV}$ and $10^{19}\,\text{GeV}$.

\medskip

To study inflation, we perform a Weyl rescaling to the Einstein frame,
\begin{equation}
  g_{\mu\nu} \;\to\; \tilde{g}_{\mu\nu}
  \;=\; \Omega^2(\phi)\, g_{\mu\nu},
  \qquad
  \Omega^2(\phi) \;=\; 1 + \xi\,\frac{\phi^2}{M_{\rm Pl}^2},
\end{equation}
which yields a canonical Einstein--Hilbert term and an effective scalar potential
\begin{equation}
  U(\phi) \;=\; 
  \frac{\lambda(\phi)}{4}\,
  \frac{\phi^4}{\left(1 + \xi \,\phi^2/M_{\rm Pl}^2\right)^2}.
\label{eq:Einstein_potential}
\end{equation}
Due to the non-minimal coupling, the kinetic term of $\phi$ becomes non-canonical in the Einstein frame. Introducing a canonically normalized field $\chi$ via
\begin{equation}
  \frac{d\chi}{d\phi}
  \;=\;
  \sqrt{
    \frac{
      1 + \xi (1+6\xi)\,\phi^2 / M_{\rm Pl}^2
    }{
      \bigl(1 + \xi\,\phi^2/M_{\rm Pl}^2\bigr)^2
    }
  },
\label{eq:chi_prime}
\end{equation}
we can express the slow-roll dynamics in terms of the Einstein-frame potential $U$ as a function of $\chi$.

\medskip

In the slow-roll approximation, the relevant parameters are
\begin{align}
  \epsilon(\chi)
  &\;=\;
  \frac{M_{\rm Pl}^2}{2}
  \left(
    \frac{1}{U}\,\frac{dU}{d\chi}
  \right)^2,
  \label{eq:epsilon_def}
  \\
  \eta(\chi)
  &\;=\;
  M_{\rm Pl}^2
  \frac{1}{U}\,\frac{d^2 U}{d\chi^2}.
  \label{eq:eta_def}
\end{align}
For numerical convenience we evaluate the derivatives with respect to $\phi$ and use the chain rule. Defining $U'(\phi) \equiv dU/d\phi$ and $\chi'(\phi) \equiv d\chi/d\phi$, we have
\begin{align}
  \frac{dU}{d\chi}
  &= \frac{U'(\phi)}{\chi'(\phi)},
  \\
  \frac{d^2U}{d\chi^2}
  &= \frac{1}{\chi'(\phi)}
     \frac{d}{d\phi}
     \left(\frac{dU}{d\chi}\right)
   \;=\;
   \frac{1}{\chi'(\phi)}
   \frac{d}{d\phi}
   \left(\frac{U'(\phi)}{\chi'(\phi)}\right).
\end{align}

The end of inflation is determined by the condition
\begin{equation}
  \epsilon(\phi_{\rm end}) \;=\; 1,
\end{equation}
which implicitly defines $\phi_{\rm end}$ (and the corresponding $\chi_{\rm end}$). The number of e-folds between a generic field value $\phi$ and the end of inflation is given by
\begin{equation}
  N(\phi)
  \;=\;
  \int_{\phi_{\rm end}}^{\phi}
    \frac{U(\varphi)}{U'(\varphi)}\,
    \frac{\chi'(\varphi)^2}{M_{\rm Pl}^2}\,
    d\varphi,
\label{eq:N_def}
\end{equation}
which we evaluate numerically using a trapezoidal integration over the $\phi$ grid. For a given target value $N_\star$ (typically $N_\star \simeq 55$), the field value at horizon exit, $\phi_\star$, is obtained by inverting Eq.~\eqref{eq:N_def} numerically.

\medskip

The inflationary observables are then computed at $\phi_\star$. The amplitude of scalar perturbations reads
\begin{equation}
  A_s
  \;=\;
  \frac{U(\phi_\star)}{24\pi^2 M_{\rm Pl}^4\, \epsilon(\phi_\star)},
  \label{eq:As_def}
\end{equation}
while the scalar spectral index and the tensor-to-scalar ratio are given by
\begin{align}
  n_s
  &= 1 - 6\,\epsilon(\phi_\star) + 2\,\eta(\phi_\star),
  \label{eq:ns_def}
  \\
  r
  &= 16\,\epsilon(\phi_\star).
  \label{eq:r_def}
\end{align}
In our analysis, for each choice of $(n, y_D, M_D, M_N, y_N)$ we first determine the RG trajectory of $\lambda(\mu)$, construct the Einstein-frame potential $U(\phi)$, and then solve for the value of the non-minimal coupling $\xi$ that reproduces the observed amplitude $A_s \simeq 2.1\times 10^{-9}$ at $N_\star = 55$. The non-minimal coupling $\xi$ is determined by requiring that the scalar
amplitude matches the observed value given in Eq.(\ref{eq:As_def}).
This condition is implemented by scanning $\xi$ over a logarithmic range and
identifying the value that satisfies the amplitude constraint.
Once $\xi$ is fixed, the inflationary observables $n_s$ and $r$ are evaluated at
$\phi_\ast(N)$, corresponding to the number of e-folds in the range
$N \in [50,60]$.
The resulting predictions in the $(n_s,r)$ plane are then compared with the
current constraints from Planck, WMAP, and BICEP/Keck \cite{Ade:2021}.

\subsection{Inflationary Predictions in the $SM+(n)VLQ+RHN$ Framework }
\begin{figure*}[h!]
    \centering

    %--- panel (a): M_D = 1.5 TeV ---%
    \begin{minipage}[t]{0.7\textwidth}
        \centering
        \includegraphics[width=\linewidth]{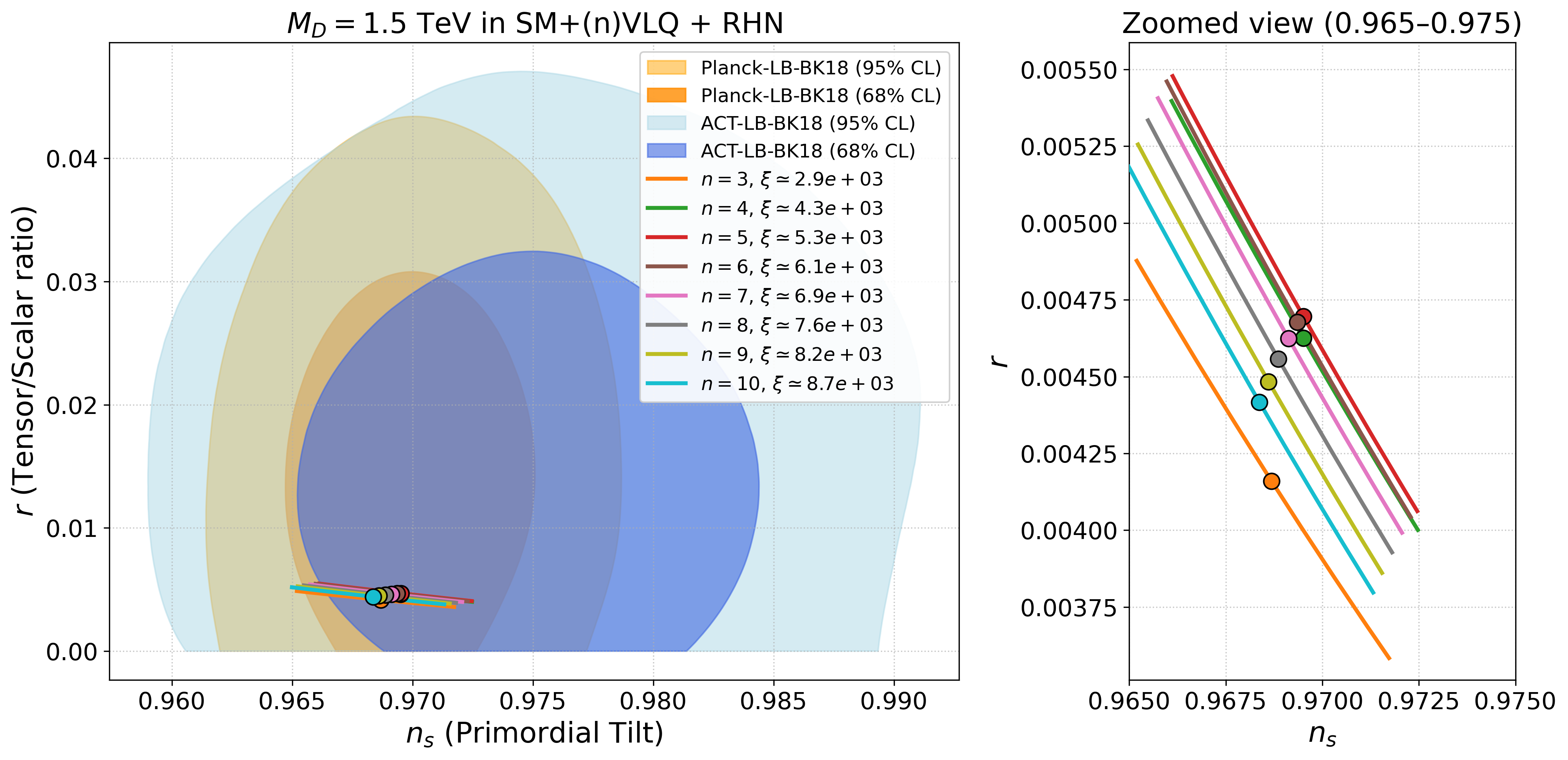}
        \\[2pt]
    \end{minipage}
    \hfill
    %--- panel (b): M_D = 3.0 TeV ---%
    \begin{minipage}[t]{0.7\textwidth}
        \centering
        \includegraphics[width=\linewidth]{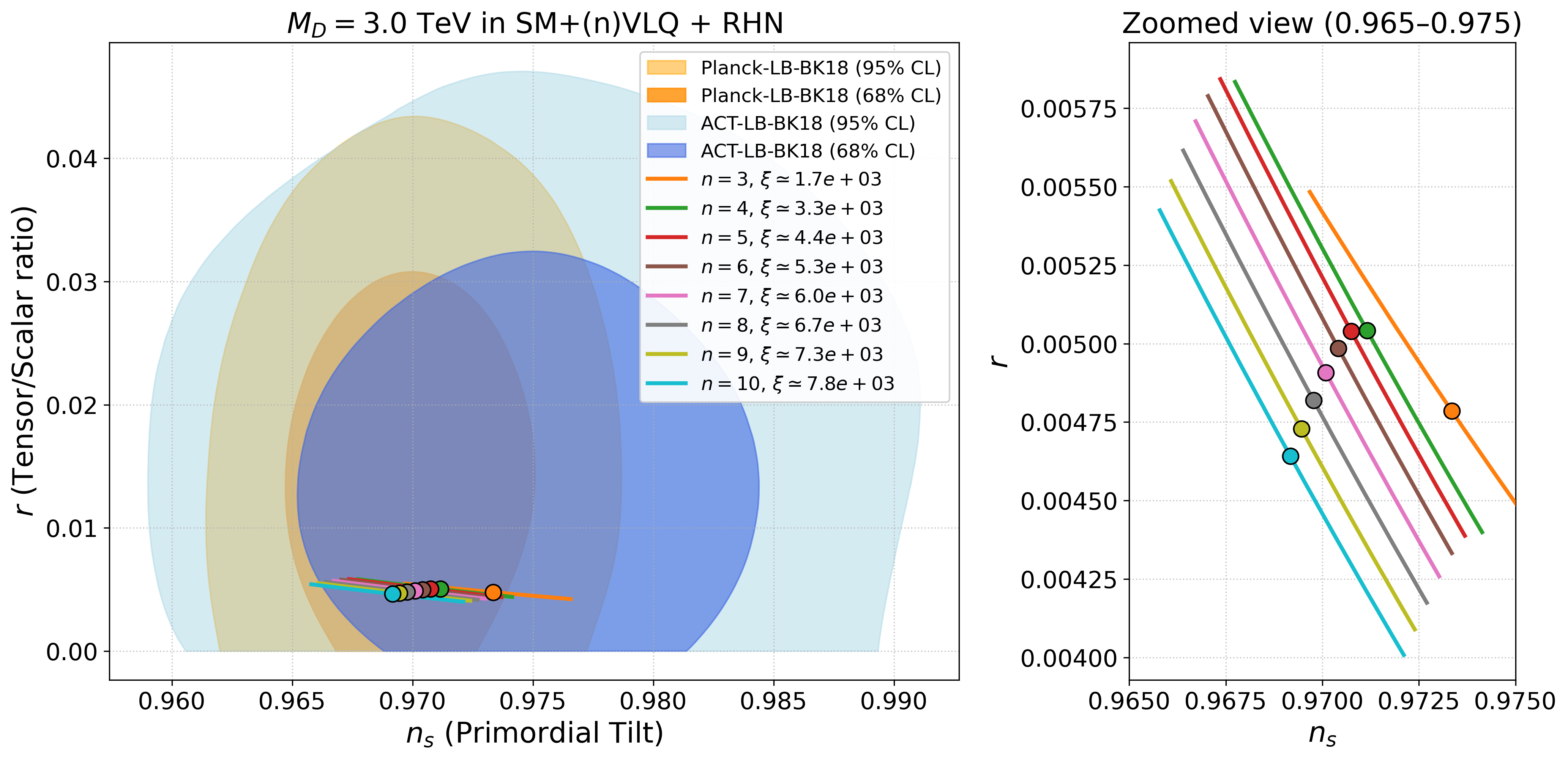}
        \\[2pt]
    \end{minipage}

    \vspace{10pt}

    %--- panel (c): M_D = 5.0 TeV ---%
    \begin{minipage}[t]{0.7\textwidth}
        \centering
        \includegraphics[width=\linewidth]{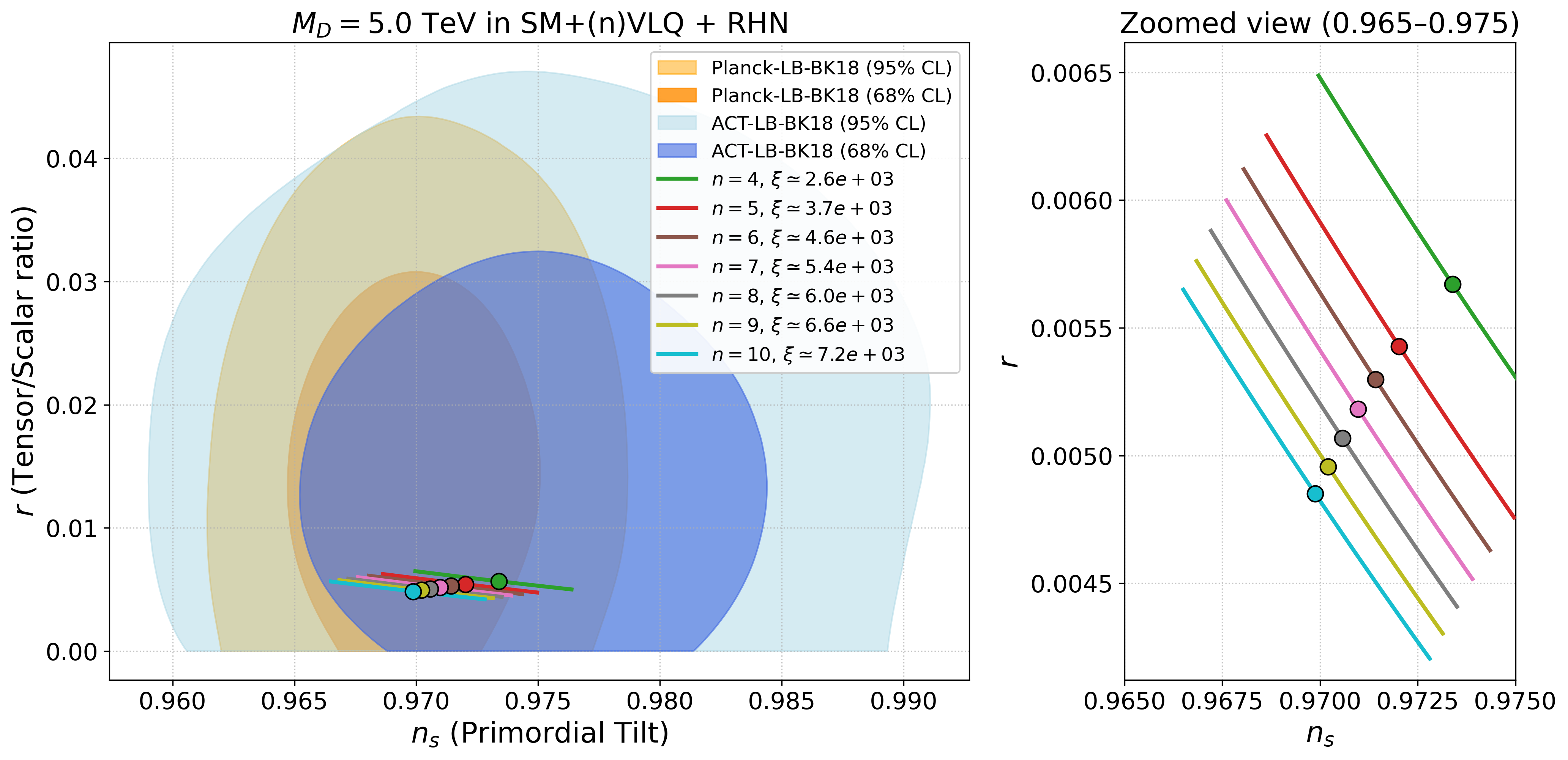}
        \\[2pt]
    \end{minipage}

    \caption{  Predictions for the spectral index $n_s$ and tensor-to-scalar ratio $r$ 
    in the SM+($n$)VLQ+RHN framework for benchmark VLQ masses 
    $M_\mathcal{D} = 1.5$, $3.0$, and $5.0~\text{TeV}$.  
    Each coloured point corresponds to a different value of $n$, with the 
    associated non-minimal coupling $\xi$ determined by the CMB amplitude.  
    The dark and light shaded regions denote the 68\% and 95\% confidence contours from CMB data, 
where $r$ is mainly constrained by BICEP/Keck (BK18) \cite{Ade:2021}, 
while $n_s$ is determined by Planck \cite{Planck:2018,Planck:2018lbu} and ACT \cite{Calabrese:2025act}, 
including CMB lensing and BAO information.
    }
    \label{fig:ns_r_RHN_three_panels}
\end{figure*}

We first present the inflationary predictions of the SM extended with $n$
degenerate down-type isosinglet VLQs and a right-handed neutrino (RHN) in the $(n_s,r)$ plane for
three benchmark VLQ masses,
$M_\mathcal{D} = 1.5,\;3.0,\;5.0~\text{TeV}$, as shown in
Fig.(~\ref{fig:ns_r_RHN_three_panels}). Throughout this analysis, we fix the benchmark Yukawa couplings to
$y_D = 0.15$ and $y_N = 0.42$, consistent with the values used in the previous section.
For each mass choice, we scan over the number of VLQs $n$ and determine, for
every $(n,M_\mathcal{D})$, the value of the non-minimal coupling $\xi$ that reproduces the
observed scalar amplitude
$A_s \simeq 2.1\times 10^{-9}$ at
$N_\ast \simeq 55$ e-folds before the end of inflation.
The resulting trajectories in the $(n_s,r)$ plane exhibit two important
features.\\
First, once the RHN is included, the points corresponding to different $n$ and
to the three benchmark values of $M_D$ cluster very close to each other. In
other words, the $(n_s,r)$ predictions become only mildly sensitive to the
number of VLQs and to their common mass. This behavior is a direct consequence
of the RG-improved Higgs potential in the inflationary regime. In the
$SM+(n)VLQ+RHN$ model, the combined effect of the RHN Yukawa coupling and the VLQ
sector renders the running of the Higgs quartic coupling $\lambda(\mu)$ in the
inflationary domain $10^{15}\!-\!10^{19}~\text{GeV}$ remarkably smooth: 
$\lambda(\mu)$ remains positive and its scale dependence is significantly
suppressed. Since the slow-roll parameters $\epsilon$ and $\eta$ are determined
by derivatives of the Einstein-frame potential
Eq.(\ref{eq:Einstein_potential}),
the small and weakly $n$-dependent slope of $\lambda(\mu)$ in this region
translates into very similar values of $\epsilon_\ast$ and $\eta_\ast$ for
different $(n,M_\mathcal{D})$. The residual dependence on $n$ and $M_D$ is further
absorbed by the $A_s$ matching condition through a mild adjustment of $\xi$,
which explains why the $n_s$–$r$ curves for the three benchmark masses lie
close to one another.

\FloatBarrier

Second, the predicted $(n_s, r)$ values for all considered $M_D$ scales (1.5, 3.0, and 5.0 TeV) lie remarkably well within the 68\% confidence level (CL) regions favored by the latest combined cosmic microwave background (CMB) constraints. As shown in the figures, the theoretical predictions are fully compatible with the joint data from Planck 2018, CMB Lensing, BAO, and BICEP/Keck 2018 (Planck-LB-BK18) \cite{Ade:2021,Planck:2018,Planck:2018lbu}. Moreover, when confronting the model with the recent Atacama Cosmology Telescope (ACT) DR6 data \cite{Calabrese:2025act}, we observe that the predictions remain comfortably within the 68\% CL of the ACT-LB-BK18 contours, perfectly accommodating the slight shift toward higher $n_s$ values favored by ACT. All viable configurations exhibit a clear tendency toward the low-$r$ regime ($r \sim 0.004-0.007$). This reflects the fact that the $SM+(n)VLQ+RHN$ setup not only stabilizes the Higgs potential up to the Planck scale but also generates an inflationary plateau sufficiently flat to yield tensor-to-scalar ratios and scalar tilts in excellent agreement with the most stringent current CMB observations.

%Second, almost all points lie well  within the region favored by the latest combined
%Planck, WMAP, and BICEP/Keck bounds \cite{Ade:2021}, with a clear tendency toward the low-$r$ regime. This reflects the fact that the
%$SM+(n)VLQ+RHN$ setup not only stabilizes the Higgs potential up to the Planck scale
%but also generates an inflationary plateau that is sufficiently flat to produce
%low tensor-to-scalar ratios and scalar tilts compatible with current CMB data.
%In summary, the RHN-induced smoothing of the RG-improved Higgs quartic in the
%inflationary range, together with the stabilizing VLQ contributions, leads to a
%set of $n_s$–$r$ predictions that are both theoretically well controlled and in
%excellent agreement with cosmological observations for all three benchmark
%values of $M_\mathcal{D}$.

\subsection{Inflationary Predictions in the $SM+(n)VLQ$ Framework }
\begin{figure*}[h!]
    \centering

    \includegraphics[width=0.5\textwidth]{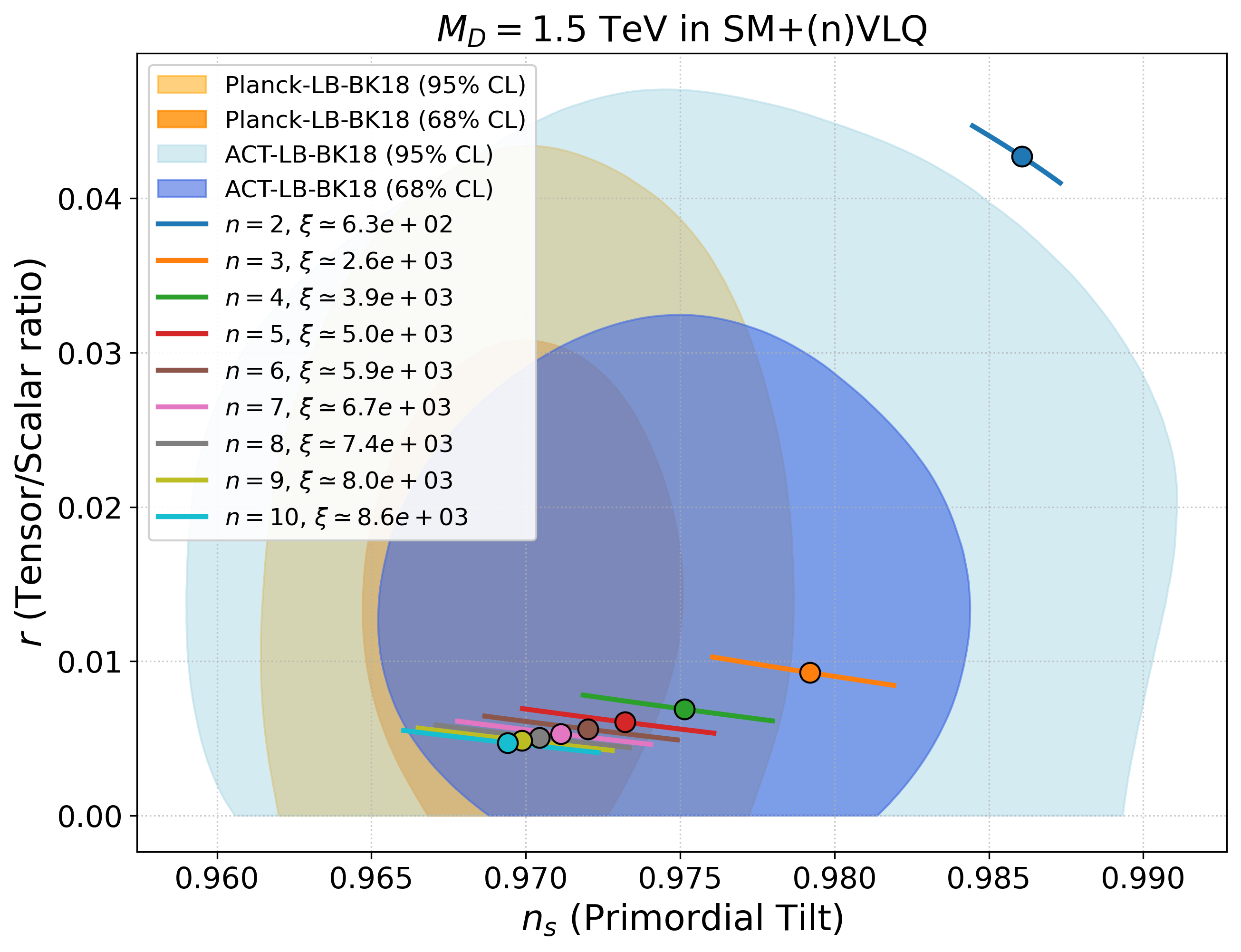}\\[-2pt]

    \vspace{6pt}

    \includegraphics[width=0.5\textwidth]{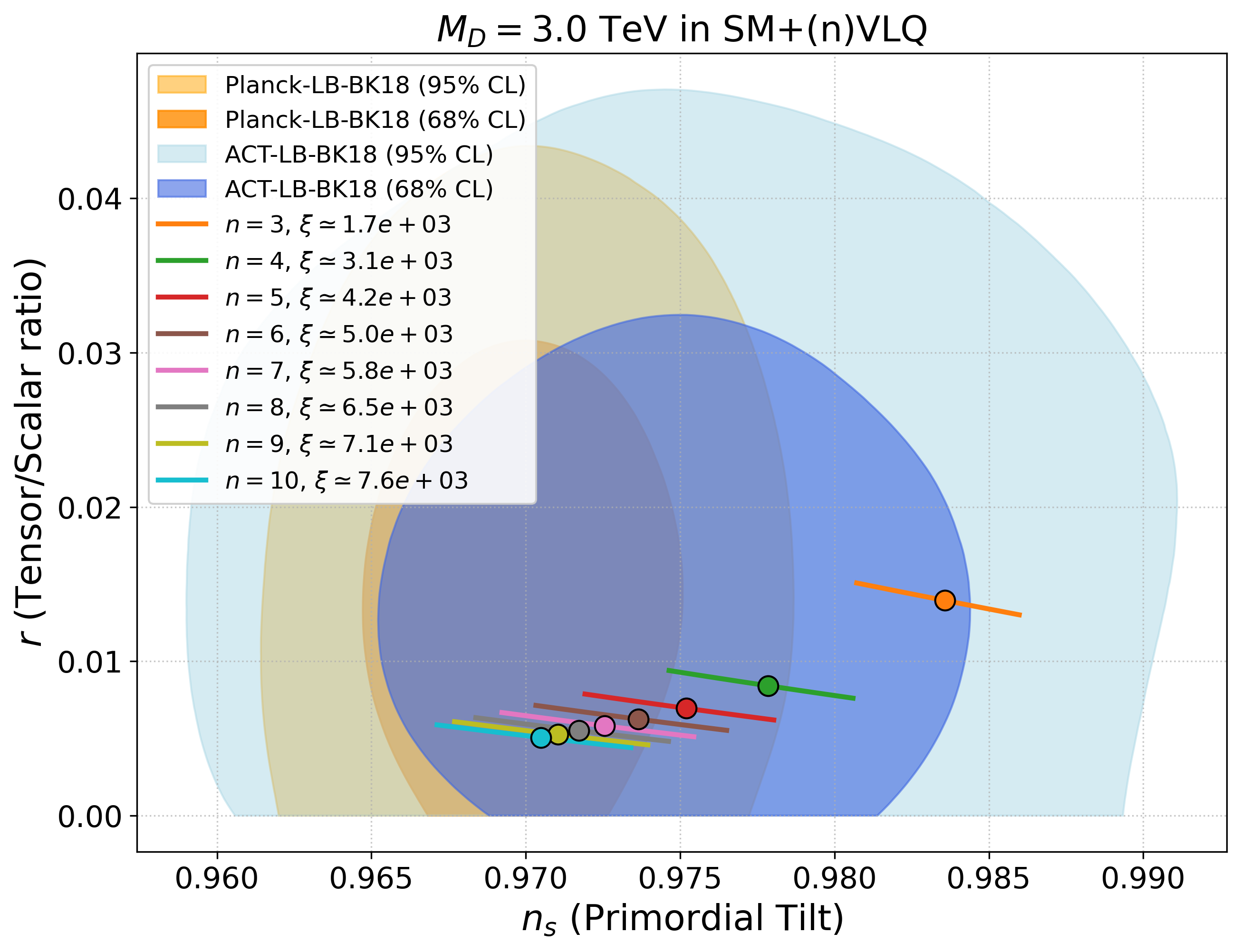}\\[-2pt]

    \vspace{6pt}

    \includegraphics[width=0.5\textwidth]{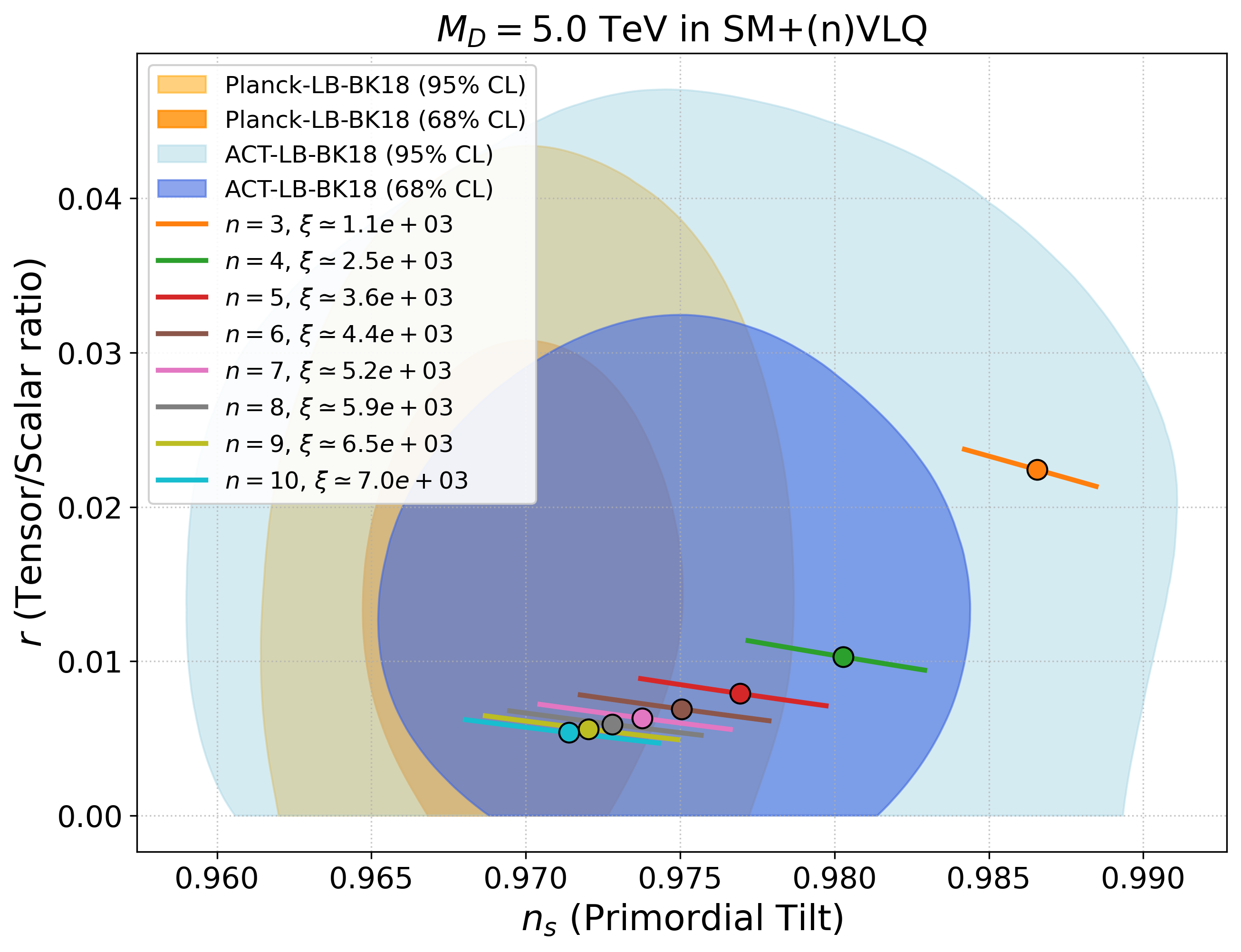}\\[-2pt]

    \caption{
    Predictions for the spectral index $n_s$ and tensor-to-scalar ratio $r$
    in the SM+($n$)VLQ framework for the benchmark VLQ masses
    $M_{\mathcal D}=1.5$, $3.0$, and $5.0~\mathrm{TeV}$.
    Each coloured point corresponds to a different value of $n$, while the
    associated non-minimal coupling $\xi$ is fixed by the CMB amplitude.
    The dark and light shaded regions denote the 68\% and 95\% confidence contours from CMB data, 
where $r$ is mainly constrained by BICEP/Keck (BK18) \cite{Ade:2021}, 
while $n_s$ is determined by Planck \cite{Planck:2018, Planck:2018lbu} and ACT \cite{Calabrese:2025act}, 
including CMB lensing and BAO information.
    }
    \label{fig:ns_r_three_panels}
\end{figure*}

Fig.~\ref{fig:ns_r_three_panels} shows the predictions in the $(n_s,r)$ plane for the SM+$(n)$VLQ model
for three benchmark VLQ masses,
$M_D = 1.5,\;3.0,\;5.0~\mathrm{TeV}$.
For each mass choice, we scan over the VLQ multiplicity $n$ and determine,
for every $(n,M_D)$, the value of the non-minimal coupling $\xi$
that reproduces the observed scalar amplitude
$A_s \simeq 2.1\times10^{-9}$ at
$N_\ast \simeq 55$ e-folds before the end of inflation.
The resulting points are overlaid on the Planck-LB-BK18 and ACT-LB-BK18 $68\%$ and $95\%$ confidence contours.\\

In contrast to the SM+$(n)$VLQ+RHN case, the SM+$(n)$VLQ predictions exhibit a 
stronger dependence on both the number of VLQs $n$ and the VLQ mass scale
$M_{\mathcal{D}}$.
This behavior reflects the fact that, in the absence of the RHN, the running of the Higgs quartic coupling $\lambda(\mu)$ in the inflationary regime,
$10^{15} -10^{19}~\mathrm{GeV}$, displays a stronger scale dependence, driven solely by the VLQ sector.
As a consequence, the slow-roll parameters $\epsilon_\ast$ and $\eta_\ast$,
which depend on derivatives of the RG-improved Einstein-frame potential Eq.(\ref{eq:Einstein_potential}),
vary more noticeably with $(n,M_{\mathcal{D}})$, leading to a broader spread of the
corresponding $(n_s,r)$ predictions.
As illustrated in Fig. 8, while a significant fraction of the parameter space in the SM+(n)VLQ model lies within the allowed regions, the predictions are notably sensitive to the number of vector-like quarks ($n$) and their mass scale ($M_{\mathcal{D}}$). Specifically, parameter choices with smaller values of $n$, particularly at higher $M_{\mathcal{D}}$, shift the predictions towards larger values of the primordial tilt ($n_s$) and the tensor-to-scalar ratio ($r$). \\

Looking at the specific configurations in detail:
\begin{itemize}
    \item The $n=2$ case (which achieves vacuum stability only for $M_{\mathcal{D}} = 1.5$ TeV) yields predictions that are entirely excluded by the 95\% CL regions of both Planck-LB-BK18 and ACT-LB-BK18.
    \item The $n=3$ case highlights a noticeable distinction between the two CMB datasets. At $M_{\mathcal{D}} = 1.5$ TeV, it lies within the 68\% CL of ACT but is on the edge of the 95\% CL for Planck. As the mass scale increases ($M_{\mathcal{D}} \ge 3.0$ TeV), the predictions shift further to the right; while they remain within the broader 95\% CL contour of ACT-LB-BK18 (which accommodates slightly higher $n_s$ values), they fall completely outside the 95\% CL region allowed by Planck-LB-BK18.
    \item For larger multiplicities ($n \ge 4$), the predictions smoothly migrate deeper into the overlapping 68\% CL regions of both Planck and ACT, demonstrating excellent agreement with the tightest cosmological bounds across all three benchmark mass scales.
\end{itemize}

\FloatBarrier

\FloatBarrier

This strong parameter sensitivity contrasts with the SM+$(n)$VLQ+RHN scenario, where the additional RHN Yukawa contribution further smooths the running of $\lambda(\mu)$ and renders the inflationary observables only mildly sensitive to $n$ and $M_{\mathcal{D}}$. The comparison highlights the crucial role of the RHN in improving the overall agreement with CMB constraints by reducing the model dependence of the $n_s$--$r$ predictions.

%While a significant fraction of the points lies within the Planck+WMAP+BICEP/Keck allowed regions, some parameter choices—particularly for smaller values of $n$ or specific $M_{\mathcal{D}}$—lead to predictions that fall outside the Planck+WMAP+BICEP/Keck confidence regions. This contrasts with the SM+$(n)$VLQ+RHN scenario, where the additional RHN Yukawa
%contribution further smooths the running of $\lambda(\mu)$ and renders the
%inflationary observables only mildly sensitive to $n$ and $M_{\mathcal{D}}$.
%The comparison highlights the crucial role of the RHN in improving the overall
%agreement with CMB constraints by reducing the model dependence of the
%$n_s$--$r$ predictions.

\section{Conclusion}
\label{conc}

The primary goal of this work is to construct a phenomenologically minimal framework that
simultaneously accounts for neutrino mass generation, electroweak vacuum
stability, and Higgs inflation within a renormalizable extension of the
Standard Model.
To this end, we study a model containing $n$ degenerate down-type isosinglet VLQs, supplemented by a single RHN
implementing a Type-I seesaw mechanism.
The analysis is carried out using the RG-improved Higgs potential, where the
running of the Higgs quartic coupling and the relevant inflationary dynamics
are determined by solving the full set of RGEs,
including the contributions from both the VLQs and RHN sectors.

We first consider the SM+$(n)$VLQ+RHN framework, in which the RHN simultaneously
generates light neutrino masses via the Type-I seesaw mechanism and affects
the RG evolution relevant for Higgs inflation.
We find that, for suitable choices of $n$ and $M_\mathcal{D}$, the combined effect of the VLQs and the RHN stabilizes the Higgs
potential up to the Planck scale, while rendering the running of the Higgs
quartic coupling remarkably smooth in the inflationary regime.
To verify the validity of our stability predictions, we performed a comprehensive sensitivity analysis with respect to the top-quark mass ($m_t$) within its current $1\sigma$ and $2\sigma$ experimental intervals. We found that for configurations with $n \ge 4$, absolute vacuum stability is consistently maintained across the entire uncertainty range for all evaluated VLQ masses, effectively ensuring a stabilization mechanism independent of $m_t$ uncertainties. As a consequence, the inflationary predictions in the $(n_s, r)$ plane exhibit only a mild dependence on the number of VLQs, $n$, and the VLQ mass scale $M_{\mathcal{D}}$, with the resulting trajectories clustering in a narrow region. These results lie remarkably well within the 68\% CL regions favored by the latest combined CMB constraints, including the recent ACT DR6 data which accommodates the slight shift toward higher $n_s$ values. 

For comparison, we also analyze the SM+$(n)$VLQ scenario by removing the RHN contributions from the RG equations. While the VLQs alone can still stabilize the Higgs potential and lead to viable Higgs inflation, the resulting inflationary observables display a stronger sensitivity to the VLQ parameters. This behavior manifests itself as a broader spread of predictions in the $(n_s,r)$ plane. Specifically, the $n=2$ case is entirely excluded by both datasets, and the $n=3$ case exhibits tension with the Planck 95\% CL at higher mass scales despite remaining compatible with ACT. In contrast, configurations with $n \geq 4$ maintain strong consistency across the overlapping constraints of both datasets.

Overall, these results highlight the complementary roles of the VLQ and RHN sectors.
The VLQs provide the dominant stabilizing effect on the Higgs potential,
whereas the RHN plays a crucial role in smoothing the RG evolution in the
inflationary domain and reducing the model dependence of the inflationary
observables.
Consequently, the SM+$(n)$VLQ+RHN framework emerges as a theoretically
well-controlled and phenomenologically viable realization of Higgs inflation,
consistent with electroweak vacuum stability and current cosmological data.

\section*{Acknowledgment} 
This work is dedicated to the memory of Durmuş Demir, whose remarkable contributions to physics and enduring passion continue to be a source of inspiration. The author thanks Beste Korutlu Yılmaz and Ozan Sargın for valuable discussions related to electroweak vacuum stability and Higgs inflation.

\appendix
\section{Renormalization Group Equations for the SM+\,$(n)$\,VLQ\,+\,RHN Model}
\label{sec:RGE_1}

In this appendix, we summarise the one-loop and (partially) two-loop 
RGEs used in the numerical analysis of the 
SM extended by $n$ down-type VLQs and a single 
Majorana RHN.  

The RG scale is defined as $t=\ln\mu$, and all $\beta$ functions follow
\begin{equation}
\beta_x \equiv \frac{dx}{dt}.
\end{equation}

\subsection{Standard Model (1-loop)}
\label{sec:SM_1loop}

At one loop the SM $\beta$-functions for the gauge couplings, top Yukawa 
and Higgs quartic coupling read
\begin{align}
\beta_{g_1}^{\rm SM,1L} &= \frac{41}{10} g_1^3, \\
\beta_{g_2}^{\rm SM,1L} &= -\frac{19}{6} g_2^3, \\
\beta_{g_3}^{\rm SM,1L} &= -7 g_3^3, \\[4pt]
\beta_{y_t}^{\rm SM,1L} &=
y_t\!\left(\frac{9}{2}y_t^2 - 8g_3^2 - \frac{9}{4}g_2^2 - \frac{17}{20}g_1^2\right), \\[4pt]
\beta_{\lambda}^{\rm SM,1L} &= 
24\lambda^2 - 6y_t^4 
+ \frac{9}{8}g_2^4 + \frac{27}{200}g_1^4 + \frac{9}{20}g_2^2 g_1^2 + \lambda\!\left(12y_t^2 - 9g_2^2 - \frac{9}{5}g_1^2\right).
\end{align}

\subsection{Standard Model (2-loop top-sector terms)}
\label{sec:SM_2loop}
We also include the dominant two-loop terms involving the top Yukawa, 
matching exactly the structure used in the numerical implementation

\begin{align}
\beta_{g_1}^{\rm SM,2L} &= g_1^3\!\left(
\frac{199}{50}g_1^2 + \frac{27}{10}g_2^2 + \frac{44}{5}g_3^2 
- \frac{17}{10}y_t^2
\right), \\[4pt]
\beta_{g_2}^{\rm SM,2L}  &= g_2^3\!\left(
\frac{9}{10}g_1^2 + \frac{35}{6}g_2^2 + 12 g_3^2 - \frac{3}{2}y_t^2
\right), \\[4pt]
\beta_{g_3}^{\rm SM,2L}  &= g_3^3\!\left(
\frac{11}{10}g_1^2 + \frac{9}{2}g_2^2 - 26 g_3^2 - 2 y_t^2
\right), \\[4pt]
\beta_{y_t}^{\rm SM,2L}  &= 
-12y_t^5 
+ y_t^3\!\left(\frac{131}{16} g_2^2 + \frac{39}{80}g_1^2 + 15g_3^2\right) \\ \nonumber
 &+ y_t\!\left(
\frac{3561}{15000}g_1^4 - \frac{27}{100}g_1^2 g_2^2 - \frac{23}{4}g_2^4
- \frac{57}{75}g_1^2 g_3^2 - 9 g_2^2 g_3^2 - 108 g_3^4
\right) \nonumber\\
& + y_t\!\left(
\frac{3}{2}\lambda^2 - 6\lambda y_t^2 + \lambda\left(3g_2^2 + \frac{9}{25}g_1^2\right)
\right), \\[4pt]
\beta_{\lambda}^{\rm SM,2L}  &= 
-312\lambda^3 
+ 36\lambda^2(3g_2^2 + \frac{3}{5}g_1^2)
- \lambda\!\left(
\frac{73}{8}g_2^4 - \frac{351}{100}g_1^2 g_2^2 - \frac{16983}{5000}g_1^4
\right) \\ \nonumber
 &+ \frac{305}{8}g_2^6 - \frac{867}{200}g_1^2 g_2^4 
- \frac{15093}{5000}g_1^4 g_2^2 - \frac{92097}{125000}g_1^6 \nonumber\\
&
- 3\lambda y_t^4 + 30 y_t^6 
- y_t^4\!\left(\frac{8}{5}g_1^2 + 8g_2^2 + 32g_3^2\right) 
+ \lambda y_t^2\!\left(\frac{45}{2}g_2^2 + \frac{17}{2}g_1^2 + 80g_3^2 - 144 y_t^2\right).
\end{align}

%-----------------------------------------------------------------------
\subsection{Down-type singlet VLQ contributions (1 loop)}
\label{sec:VLQ_1loop}
In the following, we present the RGE for 
$y_{\mathcal{D}}$ together with the corresponding one-loop contributions relevant for the present analysis \cite{Gopalakrishna:2019,Arsenault:2023vlf}.
\begin{eqnarray}
\beta_{y_{\mathcal{D}}}^{\rm VLQ} &=&
y_{\mathcal{D}}\!\left[
\frac{3}{2}y_t^2
+ \left(\frac{3}{2}+3n\right)y_{\mathcal{D}}^2
- 8g_3^2 - \frac{9}{4}g_2^2 - \frac{1}{4}g_1^2
\right], \\[6 pt]
\Delta\beta_{g_1}^{\rm VLQ} &=& \frac{4}{15} n g_1^3, \\[6 pt]
\Delta\beta_{g_2}^{\rm VLQ} &=& 0, \\
\Delta\beta_{g_3}^{\rm VLQ} &=& \frac{2}{3} n g_3^3, 
\\[6 pt]
\Delta\beta_{y_t}^{\rm VLQ} &=& 
\frac{3}{2}\,n\, y_{\mathcal{D}}^2 y_t, \\[6 pt]
\Delta\beta_{\lambda}^{\rm VLQ} &=&
-6n\, y_{\mathcal{D}}^4 + 12n\, y_{\mathcal{D}}^2\,\lambda.
\end{eqnarray}
\subsection{Majorana RHN contributions (1 loop)}
\label{sec:RHN_1loop}

In the following, we present the RGE for the RHN Yukawa coupling and retain only its dominant one-loop contributions to the Higgs quartic coupling, which are sufficient to capture the impact of the RHN sector on vacuum stability. \cite{Antusch:2003kp,Antusch:2002rr}
\begin{eqnarray}
\beta_{y_N}^{\rm RHN} &=& 
y_N\!\left(
\frac{5}{2}y_N^2 + 3n\, y_{\mathcal{D}}^2 + 3y_t^2
- \frac{9}{20}g_1^2 - \frac{9}{4}g_2^2
\right), \\[6pt]
\Delta\beta_{\lambda}^{\rm RHN} &=& 
-2y_N^4 + 4y_N^2\,\lambda.
\end{eqnarray}

%-----------------------------------------------------------------------
\subsection{Full RG system in the $SM+(n)VLQ+RHN$ Framework }
\label{sec:full_RGE}

In this part, we summarize the  RGEs  for the couplings in the  $SM+(n)VLQ+RHN$ framework. Collecting all contributions, the full $\beta$-functions used in the 
numerical analysis read
\begin{align}
\beta_{g_1} &= 
\frac{1}{16\pi^2}\beta_{g_1}^{\rm SM,1L}
+ \frac{1}{(16\pi^2)^2}\beta_{g_1}^{\rm SM,2L}
+ \frac{1}{16\pi^2}\Delta\beta_{g_1}^{\rm VLQ},  \\[6pt]
\beta_{g_2} &= 
\frac{1}{16\pi^2}\beta_{g_2}^{\rm SM,1L}
+ \frac{1}{(16\pi^2)^2}\beta_{g_2}^{\rm SM,2L}, \\[6pt]
\beta_{g_3} &= 
\frac{1}{16\pi^2}\beta_{g_3}^{\rm SM,1L}
+ \frac{1}{(16\pi^2)^2}\beta_{g_3}^{\rm SM,2L}
+ \frac{1}{16\pi^2}\Delta\beta_{g_3}^{\rm VLQ}, \\[6pt]
\beta_{y_t} &= 
\frac{1}{16\pi^2}\beta_{y_t}^{\rm SM,1L}
+ \frac{1}{(16\pi^2)^2}\beta_{y_t}^{\rm SM,2L}
+ \frac{1}{16\pi^2}\Delta\beta_{y_t}^{\rm VLQ}, \\[6pt]
\beta_{\lambda} &= 
\frac{1}{16\pi^2}\beta_{\lambda}^{\rm SM,1L}
+ \frac{1}{(16\pi^2)^2}\beta_{\lambda}^{\rm SM,2L}
+ \frac{1}{16\pi^2}\Delta\beta_{\lambda}^{\rm VLQ}
+ \frac{1}{16\pi^2}\Delta\beta_{\lambda}^{\rm RHN}, \\[6pt]
\beta_{y_{\mathcal{D}}} &= 
\frac{1}{16\pi^2}\beta_{y_{\mathcal{D}}}^{\rm VLQ}, \\[6pt]
\beta_{y_N} &= 
\frac{1}{16\pi^2}\beta_{y_N}^{\rm RHN}.
\end{align}

The RGEs for the SM+$(n)$VLQ framework are obtained by removing all
right-handed-neutrino contributions from the above equations.

\section{Input Values at the Top–Quark Mass Scale}
\label{sec:InputValues}

For the numerical solution of the RGEs,  
all SM couplings are initialized at the top–quark mass scale 
$\mu = m_t$. Unless stated otherwise, we adopt the central values reported by the Particle Data Group (PDG) ~\cite{PDG2024} and widely used in vacuum–stability analyses.  
The input parameters employed throughout this work are summarized in 
Table~\ref{tab:input_parameters}.

\begin{table}[h!]
\centering
\caption{Input values for the SM parameters at 
$\mu = m_t$
, which serve as boundary conditions for the RG evolution discussed in the main text, together with the initial conditions for the RHN and VLQ parameters defined at their respective mass thresholds.}
\label{tab:input_parameters}
\begin{tabular}{l c}
\hline\hline
Parameter & Value \\ 
\hline
Higgs mass & $m_h = 125.10~\text{GeV}$ \\
Top-quark mass & $m_t = 172.56~\text{GeV}$ \\
Strong coupling & $\alpha_s(m_Z) = 0.1181$ \\
Gauge coupling $g_1$ (GUT--normalized) & $g_1(m_t) = 0.462$ \\
Gauge coupling $g_2$ & $g_2(m_t) = 0.647$ \\
Gauge coupling $g_3$ & $g_3(m_t) = 1.166$ \\
Top Yukawa coupling & $y_t(m_t) = 0.936$ \\
Higgs quartic coupling & $\lambda(m_t) = 0.126$ \\
\hline
RHN mass & $M_N = 10^{14}~\text{GeV}$ \\
RHN Yukawa coupling & $y_N(M_N) = 0.42$ \\
VLQ Dirac mass & $M_{\mathcal{D}}=1.5, 3.0, 5.0 \ TeV$ \\
VLQ Yukawa coupling & $y_{\mathcal{D}}(M_\mathcal{D})=0.15$ \\
Number of VLQs & $n = 1-10$ \\
\hline\hline
\end{tabular}
\end{table}

\section{Impact of Additional Right-Handed Neutrinos on Vacuum Stability}
\label{sec:RHNs}

As noted in Footnote 1, while the minimal framework investigated in the main text employs a single RHN, a realistic neutrino mass generation via the Type-I seesaw mechanism requires at least two RHNs. To quantify the impact of an extended RHN sector on the electroweak vacuum stability, we investigate the cases with $N_{\text{RHN}} = 1, 2,$ and $3$ degenerate RHNs, assuming a common Yukawa coupling $y_N = 0.42$ and mass $M_N = 10^{14}$ GeV.

\begin{figure*}[h]
    \centering
    % Birinci Figür: MD = 1.5 TeV
    \begin{subfigure}[b]{0.5\textwidth}
        \centering
        \includegraphics[width=\textwidth]{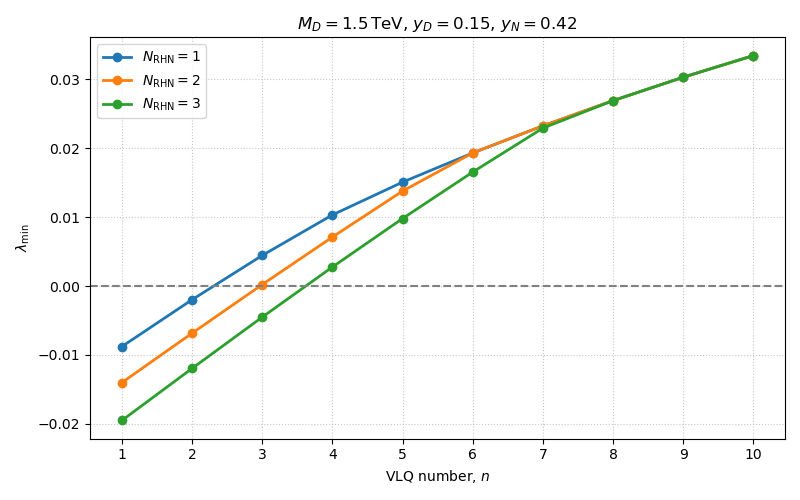}
        \caption{}
        \label{fig:md_1.5}
    \end{subfigure}
    \hfill
    % İkinci Figür: MD = 3.0 TeV
    \begin{subfigure}[b]{0.5\textwidth}
        \centering
        \includegraphics[width=\textwidth]{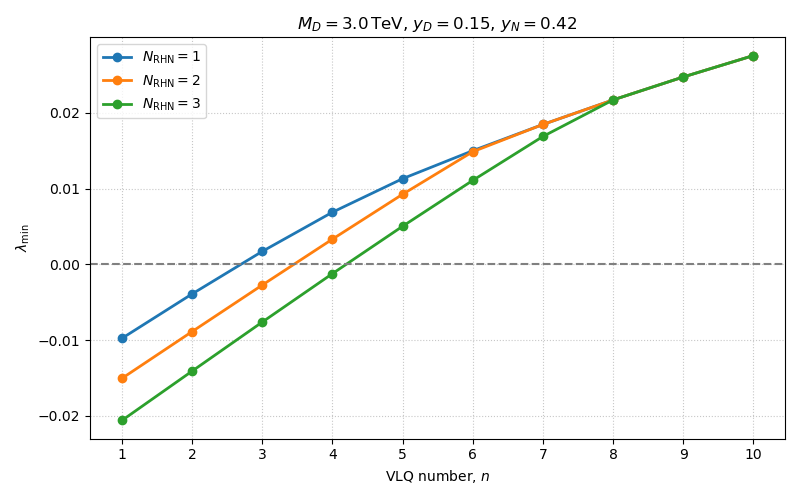}
        \caption{}
        \label{fig:md_3.0}
    \end{subfigure}
    \hfill
    % Üçüncü Figür: MD = 5.0 TeV
    \begin{subfigure}[b]{0.5\textwidth}
        \centering
        \includegraphics[width=\textwidth]{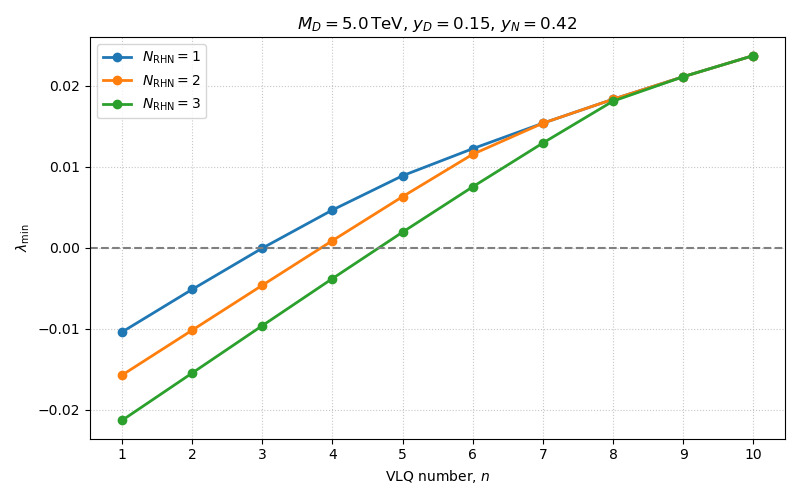}
        \caption{}
        \label{fig:md_5.0}
    \end{subfigure}
    
    \caption{The variation of the minimum quartic coupling ($\lambda_{\text{min}}$) as a function of the number of VLQs. The panels (a)-(c) correspond to the mass scales $M_D = 1.5$ TeV , $3.0$ TeV, and $5.0$ TeV, with fixed parameters $y_D = 0.15$ and $y_N = 0.42$.It should be noted that although the number of VLQs, $n$, is inherently an integer, the discrete data points in the figures are connected by continuous lines solely to guide the eye.}
    \label{fig:vacuum_stability_RHNs}
\end{figure*}

As clearly seen from the Fig.\ref{fig:vacuum_stability_RHNs} vacuum stability ($\lambda_{\text{min}} > 0$) is never achieved for $n=1$ and $n=2$ at any of the investigated mass scales ($M_D = 1.5$, $3.0$, and $5.0$ TeV), even in the most minimal scenario with a single RHN ($N_{\text{RHN}} = 1$). Driven by the increasing number of RHNs, the minimum number of VLQs required for stability shifts slightly upward (from $n=3$ to at most $n=5$, depending on the mass scale). Beyond these specific values, the stability conditions become highly robust. Regardless of the number of RHNs present ($N_{\text{RHN}} = 1$, $2$, or $3$), the conditions $n \ge 4$ for $M_{\mathcal{D}} = 1.5$ TeV and $n \ge 5$ for $M_{\mathcal{D}} = 3.0$ and $5.0$ TeV consistently ensure absolute vacuum stability.\\


\newpage


\begin{thebibliography}{9}



\bibitem{SNO2001}
Q.~R.~Ahmad \textit{et al.} [SNO Collaboration],
``Measurement of the Rate of $\nu_e + d \rightarrow p + p + e^-$ Interactions Produced by $^{8}$B Solar Neutrinos at the Sudbury Neutrino Observatory,''
Phys.\ Rev.\ Lett.\ \textbf{87}, 071301 (2001),
[arXiv:\href{https://arxiv.org/abs/nucl-ex/0106015}{nucl-ex/0106015}].

\bibitem{SuperK1998}
Y.~Fukuda \textit{et al.} [Super-Kamiokande Collaboration],
“Evidence for Oscillation of Atmospheric Neutrinos,”
Phys.\ Rev.\ Lett.\ \textbf{81}, 1562--1567 (1998),
[arXiv:\href{https://arxiv.org/abs/hep-ex/9807003}{hep-ex/9807003}].

\bibitem{Degrassi2012}
G.~Degrassi et al.,
“Higgs mass and vacuum stability in the Standard Model at NNLO,”
JHEP {\bf 08}, 098 (2012),
[arXiv:\href{https://arxiv.org/abs/1205.6497}{1205.6497} [hep-ph]]


\bibitem{Buttazzo2013}
D.~Buttazzo et al.,
“Investigating the near-criticality of the Higgs boson,”
JHEP {\bf 12}, 089 (2013),
[arXiv:\href{https://arxiv.org/abs/1307.3536}{1307.3536} [hep-ph]].

\bibitem{Starobinsky1980}
A.~A.~Starobinsky,
“A New Type of Isotropic Cosmological Models Without Singularity,”
Phys.\ Lett.\ B {\bf 91}, 99 (1980).
doi:\href{https://doi.org/10.1016/0370-2693(80)90670-X}{10.1016/0370-2693(80)90670-X}.

\bibitem{Bezrukov2008}
F.~L.~Bezrukov and M.~Shaposhnikov,
“The Standard Model Higgs boson as the inflaton,”
Phys.\ Lett.\ B {\bf 659}, 703 (2008),
[arXiv:\href{https://arxiv.org/abs/0710.3755}{0710.3755} [hep-th]].

\bibitem{Minkowski1977}
P.~Minkowski,
“$\mu \to e\gamma$ at a Rate of One Out of $10^{9}$ Muon Decays?”,
Phys.\ Lett.\ B \textbf{67}, 421–428 (1977),
doi:\href{https://doi.org/10.1016/0370-2693(77)90435-X}{10.1016/0370-2693(77)90435-X}.

\bibitem{GellMann1979}
M.~Gell-Mann, P.~Ramond, and R.~Slansky,
“Complex Spinors and Unified Theories,”
in \textit{Supergravity}, 
edited by P.~van Nieuwenhuizen and D.~Z.~Freedman 
(North Holland, Amsterdam, 1979), p.~315
[arXiv:\href{https://arxiv.org/abs/1306.4669}{1306.4669} [hep-th]].

\bibitem{Yanagida1979}
T.~Yanagida,
in \textit{Proceedings of the Workshop on Unified Theory and the Baryon Number of the Universe}, 
edited by O.~Sawada and A.~Sugamoto 
(KEK, Tsukuba, 1979).

\bibitem{MohapatraSenjanovic1980}
R.~N.~Mohapatra and G.~Senjanović,
“Neutrino Mass and Spontaneous Parity Nonconservation,”
Phys.\ Rev.\ Lett.\ \textbf{44}, 912 (1980),
doi:\href{https://doi.org/10.1103/PhysRevLett.44.912}{10.1103/PhysRevLett.44.912}.

\bibitem{Cheng:1980}
T.~P.~Cheng and L.~F.~Li,
``Neutrino Masses, Mixings and Oscillations in SU(2) \(\times\) U(1) Models of Electroweak Interactions,''
Phys.\ Rev.\ D \textbf{22}, 2860 (1980),
doi:\href{https://doi.org/10.1103/PhysRevD.22.2860}{10.1103/PhysRevD.22.2860}.


\bibitem{Mohapatra:1981}
R.~N.~Mohapatra and G.~Senjanovi\'c,
``Neutrino masses and mixings in gauge models with spontaneous parity violation,''
Phys.\ Rev.\ D \textbf{23}, 165 (1981),
doi:\href{https://doi.org/10.1103/PhysRevD.23.165}{10.1103/PhysRevD.23.165}.


\bibitem{Foot:1989}
R.~Foot, H.~Lew, X.~G.~He and G.~C.~Joshi,
``Seesaw Neutrino Masses Induced by a Triplet of Leptons,''
Z.\ Phys.\ C \textbf{44}, 441–444 (1989),
doi:\href{https://doi.org/10.1007/BF01415558}{10.1007/BF01415558}.


\bibitem{Demir:2021}
D.~Demir, C.~Karahan and O.~Sarg{\i}n,
``Type-3/2 seesaw mechanism,''
Phys. Rev. D \textbf{104}, no.7, 075038 (2021),
[arXiv:\href{https://arxiv.org/abs/2105.06539}{2105.06539} [hep-ph]].


\bibitem{Sher1989}
M.~Sher,
“Electroweak Higgs Potentials and Vacuum Stability,”
Phys.\ Rept.\ \textbf{179}, 273–418 (1989),
doi:\href{https://doi.org/10.1016/0370-1573(89)90061-6}{10.1016/0370-1573(89)90061-6}.


\bibitem{Chetyrkin:2012rz}
K.~G.~Chetyrkin and M.~F.~Zoller,
``Three-loop beta functions for the Higgs self-interaction and the
top-Yukawa coupling,''
JHEP \textbf{06}, 033 (2012),
[arXiv:\href{https://arxiv.org/abs/1205.2892}{1205.2892} [hep-ph]].


\bibitem{Gonderinger:0910.3167}
M.~Gonderinger, Y.~Li, H.~Patel and M.~J.~Ramsey-Musolf,
``Vacuum Stability, Perturbativity, and Scalar Singlet Dark Matter,''
JHEP \textbf{01}, 053 (2010),
[arXiv:\href{https://arxiv.org/abs/0910.3167}{0910.3167} [hep-ph]].

\bibitem{Ghorbani:2104.09542}
P.~Ghorbani,
``Vacuum stability vs. positivity in real singlet scalar extension of the standard model,''
Nucl. Phys. B \textbf{971}, 115533 (2021),
[arXiv:\href{https://arxiv.org/abs/2104.09542}{2104.09542} [hep-ph]].

\bibitem{Peli:2022}
Z.~Péli and Z.~Trócsányi,
``Vacuum stability and scalar masses in the superweak extension of the Standard Model,''
Phys.\ Rev.\ D \textbf{106}, 055045 (2022),
[arXiv:\href{https://arxiv.org/abs/2204.07100}{2204.07100} [hep-ph]].

\bibitem{Hiller:2024}
G.~Hiller, T.~Höhne, D.~F.~Litim and T.~Steudtner,
``Vacuum Stability in the Standard Model and Beyond,''
Phys.\ Rev.\ D \textbf{110}, 115017 (2024),
[arXiv:\href{https://arxiv.org/abs/2401.08811}{2401.08811} [hep-ph]]

\bibitem{Rodejohann:2012}
W.~Rodejohann and H.~Zhang,
``Impact of massive neutrinos on the Higgs self-coupling and electroweak vacuum stability,''
JHEP \textbf{06}, 022 (2012),
[arXiv:\href{https://arxiv.org/abs/1203.3825}{1203.3825} [hep-ph]].


\bibitem{Datta:2013}
A.~Datta, A.~Elsayed, S.~Khalil and A.~M.~Moretti,
``Higgs vacuum stability in the $B-L$ extended Standard Model,''
Phys.\ Rev.\ D \textbf{88}, 053011 (2013),
[arXiv:\href{https://arxiv.org/abs/1308.0816}{1308.0816} [hep-ph]].


\bibitem{Oda:2015}
S.~Oda, N.~Okada and D.~s.~Takahashi,
`Classically conformal U(1)' extended standard model and Higgs vacuum stability,''
Phys. Rev. D \textbf{92}, no.1, 015026 (2015),
[arXiv:\href{https://arxiv.org/abs/1504.06291}{1504.06291} [hep-ph]].


\bibitem{Gopalakrishna:2019}
S.~Gopalakrishna and A.~Velusamy,
``Higgs vacuum stability with vectorlike fermions,''
Phys.\ Rev.\ D \textbf{99}, 115020 (2019),
[arXiv:\href{https://arxiv.org/abs/1812.11303}{1812.11303} [hep-ph]].


\bibitem{Arsenault:2023vlf}
A.~Arsenault, K.~Y.~Cingiloglu and M.~Frank,
``Vacuum stability in the Standard Model with vector-like fermions,''
Phys.\ Rev.\ D \textbf{107}, 036018 (2023),
[arXiv:\href{https://arxiv.org/abs/2207.10332}{2207.10332} [hep-ph]].

\bibitem{referee_1}
H.~Y.~Chen, I.~Gogoladze, S.~Hu, T.~Li and L.~Wu,
%``The Minimal GUT with Inflaton and Dark Matter Unification,''
Eur. Phys. J. C \textbf{78}, no.1, 26 (2018)
doi:10.1140/epjc/s10052-017-5496-z
[arXiv:1703.07542 [hep-ph]].

\bibitem{referee_2}
N.~Okada, S.~Okada and D.~Raut,
%``SU(5)$\times$U(1)$_X$ grand unification with minimal seesaw and $Z^\prime$-portal dark matter,''
Phys. Lett. B \textbf{780}, 422-426 (2018)
doi:10.1016/j.physletb.2018.03.031
[arXiv:1712.05290 [hep-ph]].

\bibitem{Lerner:2009xg}
R.~N.~Lerner and J.~McDonald,
``Gauge singlet scalar as inflaton and thermal relic dark matter,''
Phys.\ Rev.\ D \textbf{80}, 123507 (2009),
[arXiv:\href{https://arxiv.org/abs/0909.0520}{0909.0520} [hep-ph]].


\bibitem{Nakayama:2010}
K.~Nakayama and F.~Takahashi,
``Running kinetic inflation,''
JCAP \textbf{1011}, 009 (2010),
[arXiv:\href{https://arxiv.org/abs/1008.2956}{1008.2956} [hep-ph]].



\bibitem{Barvinsky:2008}
A.~O.~Barvinsky, A.~Y.~Kamenshchik and A.~A.~Starobinsky,
``Inflation scenario via the Standard Model Higgs boson and LHC,''
JCAP \textbf{11}, 021 (2008),
[arXiv:\href{https://arxiv.org/abs/0809.2104}{0809.2104} [hep-ph]].

\bibitem{Bauer:2008}
F.~Bauer and D.~A.~Demir,
``Inflation with Non-Minimal Coupling: Metric versus Palatini Formulations,''
Phys. Lett. B \textbf{665}, 222-226 (2008),
[arXiv:\href{https://arxiv.org/abs/0803.2664}{0803.2664} [hep-ph]].

\bibitem{Rubio:2018}
J.~Rubio,
``Higgs inflation,''
Front. Astron. Space Sci. \textbf{5}, 50 (2019),
[arXiv:\href{https://arxiv.org/abs/1807.02376}{1807.02376} [hep-ph]].



\bibitem{PDG2023}
R.~L.~Workman \textit{et al.} (Particle Data Group),
“Review of particle physics,”
Prog.\ Theor.\ Exp.\ Phys. 2022, 083C01 (2022),
doi:\href{https://doi.org/10.1093/ptep/ptac097}{10.1093/ptep/ptac097}.

\bibitem{Salas2018} 
P.~F.~De Salas, S.~Gariazzo, O.~Mena, C.~A.~Ternes and M.~T{\'o}rtola,
``Neutrino Mass Ordering from Oscillations and Beyond: 2018 Status and Future Prospects,''
Front. Astron. Space Sci. \textbf{5}, 36 (2018),
[arXiv:\href{https://arxiv.org/abs/1806.11051}{1806.11051} [hep-ph]].





\bibitem{ATLAS:VLQpair2018}
M.~Aaboud \textit{et al.} [ATLAS],
``Search for pair production of heavy vector-like quarks decaying into high-$p_T$ $W$ bosons and top quarks in the lepton-plus-jets final state in $pp$ collisions at $\sqrt{s}=13$ TeV with the ATLAS detector,''
JHEP \textbf{08}, 048 (2018),
[arXiv:\href{https://arxiv.org/abs/1806.01762}{1806.01762} [hep-ex]].

\bibitem{CMS:2209.07327}
A.~Tumasyan \textit{et al.} [CMS],
``Search for pair production of vector-like quarks in leptonic final states in proton-proton collisions at $ \sqrt{s} $ = 13 TeV,''
JHEP \textbf{07}, 020 (2023),
[arXiv:\href{https://arxiv.org/abs/2209.07327}{2209.07327} [hep-ex]].

\bibitem{Chen:2017qev}
C.~Y.~Chen, S.~Dawson and E.~Furlan,
``Vectorlike fermions and Higgs effective field theory revisited,''
Phys. Rev. D \textbf{96}, no.1, 015006 (2017),
[arXiv:\href{https://arxiv.org/abs/1703.06134}{1703.06134} [hep-ph]].


\bibitem{AguilarSaavedra:2013qpa}
J.~A.~Aguilar-Saavedra, R.~Benbrik, S.~Heinemeyer and M.~P{\'e}rez-Victoria,
``Handbook of vectorlike quarks: Mixing and single production,''
Phys. Rev. D \textbf{88}, no.9, 094010 (2013),
[arXiv:\href{https://arxiv.org/abs/1306.0572}{1306.0572} [hep-ph]].


\bibitem{Lavoura:1992np}
L.~Lavoura and J.~P.~Silva,
``Bounds on the mixing of the down-type quarks with vectorlike singlet quarks,''
Phys.\ Rev.\ D {\bf 47}, 1117 (1993),
doi:\href{https://doi.org/10.1103/PhysRevD.47.1117}{10.1103/PhysRevD.47.1117}.



\bibitem{Adhikary:2024}
A.~Adhikary, M.~Olechowski, J.~Rosiek and M.~Ryczkowski,
``Theoretical constraints on models with vectorlike fermions,''
Phys.\ Rev.\ D \textbf{110}, 075029 (2024),
[arXiv:\href{https://arxiv.org/abs/2406.16050}{2406.16050} [hep-ph]].

\bibitem{DicusMathur1973}
D.~A.~Dicus and V.~S.~Mathur,
``Upper bounds on the values of masses in unified gauge theories,''
Phys.\ Rev.\ D {\bf 7}, 3111 (1973),
doi:\href{https://doi.org/10.1103/PhysRevD.7.3111}{10.1103/PhysRevD.7.3111}.

\bibitem{Chanowitz1978}
M.~S.~Chanowitz, M.~A.~Furman and I.~Hinchliffe,
``Weak Interactions of Ultraheavy Fermions. 2.,''
Nucl. Phys. B \textbf{153}, 402-430 (1979),
doi:\href{https://doi.org/10.1016/0550-3213(79)90606-0}{10.1016/0550-3213(79)90606-0}.

\bibitem{Bezrukov:2009}
F.~Bezrukov, A.~Magnin, M.~Shaposhnikov and S.~Sibiryakov,
``Higgs inflation: critical point and running strategy,''
JHEP \textbf{07} (2009) 089,
doi:10.1088/1126-6708/2009/07/089
[arXiv:0812.4950 [hep-ph]].

\bibitem{Hamada:2014}
Y.~Hamada, H.~Kawai, K.~y.~Oda and S.~C.~Park,
``Higgs Inflation is Still Alive after BICEP2,''
Phys. Rev. Lett. \textbf{113} (2014) no.14, 141301,
doi:10.1103/PhysRevLett.113.141301
[arXiv:1403.5043 [hep-ph]].


\bibitem{Ade:2021}
P.~A.~R.~Ade \textit{et al.} [BICEP and Keck],
``Improved Constraints on Primordial Gravitational Waves using Planck, WMAP, and BICEP/Keck Observations through the 2018 Observing Season,''
Phys. Rev. Lett. \textbf{127}, no.15, 151301 (2021),
[arXiv:\href{https://arxiv.org/abs/2110.00483}{2110.00483} [astro-ph.CO]].

\bibitem{Planck:2018}
Y.~Akrami \textit{et al.} [Planck Collaboration],
``Planck 2018 results. X. Constraints on inflation,''
Astron. Astrophys. \textbf{641} (2020), A10,
[arXiv:1807.06211].


\bibitem{Planck:2018lbu}
N.~Aghanim \textit{et al.} [Planck],
%``Planck 2018 results. VIII. Gravitational lensing,''
Astron. Astrophys. \textbf{641}, A8 (2020)
doi:10.1051/0004-6361/201833886
[arXiv:1807.06210 [astro-ph.CO]].

\bibitem{Calabrese:2025act}
E.~Calabrese \textit{et al.} [ACT Collaboration],
``The Atacama Cosmology Telescope: DR6 constraints on extended cosmological models,''
JCAP \textbf{11} (2025), 063,
[arXiv:2503.14454].


\bibitem{ChengLi} 
T. P. Cheng and L. F. Li, \emph{Gauge Theory of Elementary Particle Physics}, Oxford University Press (1984).

\bibitem{MV1983} 
M.E. Machacek and M.T. Vaughn, "Two-loop Renormalization Group Equations in a General Quantum Field Theory (I). Gauge Fields", Nucl. Phys. B222 (1983) 83,
doi:\href{https://doi.org/10.1016/0550-3213(83)90610-7}{10.1016/0550-3213(83)90610-7}.


\bibitem{Antusch:2003kp}
S.~Antusch, M.~Drees, J.~Kersten, M.~Lindner and M.~Ratz,
``Neutrino mass operator renormalization revisited,''
Phys. Lett. B \textbf{519}, 238-242 (2001),
[arXiv:\href{https://arxiv.org/abs/hep-ph/0108005}{hep-ph/0108005} [hep-ph]].

\bibitem{Antusch:2002rr}
S.~Antusch, M.~Drees, J.~Kersten, M.~Lindner and M.~Ratz,
``Neutrino mass operator renormalization in two Higgs doublet models and the MSSM,''
Phys. Lett. B \textbf{525}, 130-134 (2002),
[arXiv:\href{https://arxiv.org/abs/hep-ph/0110366}{hep-ph/0110366} [hep-ph]]

\bibitem{PDG2024}
S.~Navas \textit{et al.} (Particle Data Group),
``Review of Particle Physics,''
Phys.\ Rev.\ D \textbf{110}, 030001 (2024),
doi:\href{https://doi.org/10.1103/PhysRevD.110.030001}{10.1103/PhysRevD.110.030001}.

\end{thebibliography}
\end{document}